\documentclass[a4paper, 11pt]{article}
\pdfoutput=1

\usepackage{jcappub}
\usepackage[utf8]{inputenc}
\usepackage{graphicx}
\usepackage{amsmath}
\usepackage{amsfonts}
\usepackage{amssymb}
\usepackage{bbold}
\usepackage{float}

\usepackage{tikz}
\usetikzlibrary{arrows,backgrounds,snakes,patterns}
\usetikzlibrary{shapes,arrows,chains}
\usepackage{verbatim}
\usepackage{booktabs}
\usepackage{pgfplots}
\pgfplotsset{compat=1.10}
\usepgfplotslibrary{fillbetween}
\usepackage[flushleft]{threeparttable}

\usepackage{array}
\usepackage{placeins}

\usepackage{subcaption}
\usepackage{datetime2}

\renewcommand{\(}{\left(}
\renewcommand{\)}{\right)}
\renewcommand{\[}{\left[}
\renewcommand{\]}{\right]}

\newcommand{\cD}{{\cal D}}

\newcommand{\cO}{\mathcal O}
\newcommand{\df}{\dot{\phi}}
\newcommand{\dbf}{\dot{\bar{\phi}}}

\newcommand{\Mpl}{\ensuremath M_{\textrm{Pl}{}}}

\numberwithin{equation}{section}
\numberwithin{figure}{section}

\dedicated{UTTG-10-2020}
\title{The Multi-Field, Rapid-Turn Inflationary Solution}

\author[a]{Vikas Aragam}
\author[a]{Sonia Paban}
\author[a]{Robert Rosati}

\affiliation[a]{
	Theory Group, Department of Physics, University of Texas at Austin, Austin, TX 78712, USA
}

\emailAdd{aragam@utexas.edu}
\emailAdd{paban@physics.utexas.edu}
\emailAdd{rjrosati@utexas.edu}

\abstract{
There are well-known criteria on the potential and field-space geometry for determining if slow-roll, slow-turn, multi-field inflation is possible. However, even though it has been a topic of much recent interest, slow-roll, rapid-turn inflation only has such criteria in the restriction to two fields.
In this work, we generalize the two-field, rapid-turn inflationary attractor to an arbitrary number of fields. We quantify a limit, which we dub \textit{extreme turning}, in which rapid-turn solutions may be found efficiently and develop methods to do so. In particular, simple results arise when the covariant Hessian of the potential has an eigenvector in close alignment with the gradient -- a situation we find to be common and we prove generic in two-field hyperbolic geometries.
We verify our methods on several known rapid-turn models and search two type-IIA constructions for rapid-turn trajectories. For the first time, we are able to efficiently search for these solutions and even exclude slow-roll, rapid-turn inflation from one potential.
}
\arxivnumber{2010.15933}
\begin{document}
\maketitle
\thispagestyle{empty}
\newpage

\setcounter{page}{1}

\section{Introduction}

Rapidly turning trajectories in multi-dimensional field spaces have been of much recent interest: they can yield inflation and quintessence in potentials satisfying the de-Sitter conjecture \cite{Obied:2018sgi,Ooguri:2018wrx,Garg:2018reu,Garg:2018zdg,Denef:2018etk,Andriot:2018mav,Achucarro:2018vey,Agrawal:2018own,Brown:2017osf,Bjorkmo:2019aev,Aragam:2019khr,Cicoli:2020noz,Renaux-Petel:2015mga,Garcia-Saenz:2018ifx,Chakraborty:2019dfh,Akrami:2020zfz,Andriot:2020wpp}, and they can form primordial black holes \cite{Palma:2020ejf,Fumagalli:2020adf,Braglia:2020eai,Aldabergenov:2020bpt}. Rapidly turning inflation models are also profound in their own right, as they can exhibit dynamics very different from single-field inflation while being phenomenologically viable \cite{Christodoulidis:2018qdw,Christodoulidis:2019mkj,Christodoulidis:2019jsx,Achucarro:2019pux,Achucarro:2019mea,Iarygina:2020dwe,Garcia-Saenz:2018ifx,Garcia-Saenz:2018vqf,Chakraborty:2019dfh}.

The turning rate $ \omega $ of a classical trajectory is related through the equations of motion to the parameters $ \epsilon$, $\eta$, and $ \epsilon_V $,
\begin{equation}
 \epsilon_V= \epsilon \left\{ \left(1 + \frac{\eta}{2 (3-\epsilon)}\right)^2 + \frac{\omega^2}{9 H^2} \frac{1}{(1-\epsilon/3)^2} \right\},
 \label{eq:Ev_Eh_relationship}
\end{equation}
where the slow-roll and gradient parameters are defined as:
\begin{align}
	\epsilon \equiv -\frac{\dot{H}}{H^2}, \quad \eta \equiv \frac{\dot{\epsilon}}{H\epsilon}, \quad \epsilon_V \equiv \frac{\Mpl^2}{2} \frac{\left| \nabla V \right|^2}{V^2}.
\end{align}
Slow-roll, slow-turn inflation occurs when $\omega^2/9 H^2\ll 1$, $\eta \ll 1$, and $\epsilon \simeq \epsilon_V \ll 1$. Bounds on $\epsilon_V$, which depend only on the potential and field-space geometry, can be used to exclude this type of inflationary trajectory. In \cite{Hertzberg:2007wc,Flauger:2008ad,Caviezel:2008tf,Andriot:2020lea}, no-go theorems were shown to bound $\epsilon_V$ to be greater than $\cO(1)$ constants in several classes of potentials.

Until recently, there were no similar universal criteria for trajectories with large $\omega$ and $\epsilon_V$, but small $\epsilon$ and $\eta$.
Such criteria were only known for specific forms of potentials and field-space metrics (see, e.g. \cite{Brown:2017osf,Garcia-Saenz:2018ifx,Achucarro:2019pux}).
We designate these solutions as having slow-roll, rapid-turn inflation.
Last year, Bjorkmo \cite{Bjorkmo:2019fls} published a two-field rapid-turn inflationary attractor, which allows a straightforward calculation of $\omega$ and $\epsilon$ in terms of only potential and field-space metric parameters. This enables one to find regions of field space admitting rapid-turn inflation without evolving individual trajectories.

In this work, we generalize Bjorkmo's result to an arbitrary number of fields and develop methods for solving the resulting equations of motion across the entirety of field space (Section \ref{sec:RTA}). Similar to the two-field case, the field velocities are entirely determined by the assumptions of slow-roll and rapid-turning. However, unlike the two-field case, the magnitudes of the acceleration terms are not uniquely determined by slow-roll. We carefully analyze the relation between the accelerations and the turning rate. This leads us to define the \textit{extreme turning} limit, in which most of these accelerations are negligible. We then discover a purely algebraic, covariant matrix equation to solve for the field velocities, and introduce a numerical algorithm to solve it in full generality. 

A particularly simple solution to this equation arises when an eigenvector of the potential's covariant Hessian matrix approximately aligns with the potential's gradient vector. In Section \ref{sec:alignedHessian}, we observe that this Hessian alignment is rather common, analytically find the implied rapid-turn solution and develop a perturbation series around the alignment of the Hessian. 

We explore several potentials, including some known to have rapid-turning solutions and two type-IIA constructions in Section \ref{sec:examples}. We are able to exclude inflation in one of the type-IIA potentials, and identify short-lived rapid-turn initial conditions in another.

\section{The Rapid-Turn Solution}
\label{sec:RTA}

We work in an FLRW spacetime, with multiple scalar fields $\phi^a$ minimally coupled to gravity, a scalar potential $V(\phi)$, and an arbitrary kinetic term of the form $G_{ab}\df^a\df^b/2$. The kinetic term coefficients $G_{ab}(\phi)$ may be viewed as components of a metric on field-space. The usual equations of motion,
\begin{align}
	\cD_t \df^a + 3H\df^a + G^{ab} V_{,b} = 0,
\end{align}
do not readily reveal rapid-turn inflationary solutions: the kinematic basis $\{\df^a |a=1,\ldots,N_f \}$ singles out no particular direction that is tracked by turning trajectories without solving the equations of motion. As shown in \cite{Bjorkmo:2019aev}, a basis consisting of the potential's gradient vector proves much more useful, since turning trajectories have a non-vanishing velocity along the directions perpendicular to the gradient.

In this choice of basis, the field velocity is expanded as $\dot{\phi}^a = (\dot{\phi}_b v^b) v^a + \dot{\phi}_{b} \perp^{ab}$, where $v^a$ is a unit vector in the gradient direction of the potential. The velocity parallel to the gradient is $\df_v \equiv \df_a v^a$, and the projector $\perp^{ab} \equiv G^{ab} - v^a v^b$ gives the component of the field velocity perpendicular to the gradient, $\dot\phi_\perp^a \equiv \perp^a_b \dot\phi^b$. The norm can then be written $\dot\phi^2 = \dot\phi_v^2 + \dot\phi_\perp^2$, where $\dot\phi_\perp^2 = \dot\phi_{\perp b} \dot\phi_\perp^b$. The equations of motion in this basis are:
\begin{align}
\ddot\phi_\perp^a \equiv \cD_t(\perp^a_b \dot\phi^b) &= -3 H \dot\phi_\perp^a - \frac{\perp^{ab} V_{;bc}\dot\phi^c}{V_v} \dot\phi_v - v^a \frac{\dot{\phi}^b_\perp V_{;bc}\dot\phi^c}{V_v} \label{eq:PerpEOM}, \\
\ddot\phi_v \equiv \cD_t(\dot\phi^a v_a) &= -3 H \dot\phi_v - V_v + \frac{\dot\phi_\perp^a V_{;ab} \dot\phi^b}{V_v}, \label{eq:GradEOM}
\end{align}
where the norm of the potential gradient is given by $V_v \equiv \sqrt{G^{ab} V_{;a}V_{;b} }$, and $V_{;ab}$ denotes the covariant Hessian matrix of the potential. $G^{ab}$ is the inverse metric on field space. The velocity direction's unit vector is:
\begin{align}
\hat\sigma^a = \frac{\dot\phi_v v^a + \dot\phi_\perp^a}{\df}.
\end{align}
In analogy with the two-field result, we can define $\perp^a \equiv \dot\phi_\perp^a /\dot\phi_\perp$ so that $\hat\sigma^a = (\dot\phi_v v^a + \dot\phi_\perp \perp^a ) / \dot\phi$. The turning vector $\omega^a \equiv \cD_t\hat\sigma^a$ and its corresponding unit vector $\hat{s}^a$ then take the form:
\begin{align}
\omega^a &= \frac{\left( -\dot\phi_\perp^2 v^a + \dot\phi_\perp^a \dot\phi_v \right) V_{v}}{\df^3}, \\
\hat{s}^a &\equiv \frac{\omega^a }{\left| \omega^a \right|} =  \frac{-\dot\phi_\perp v^a + \perp^a \dot\phi_v}{\df}.
\end{align}
The advantage of this basis is now manifest, as the turning rate is nonzero whenever the trajectory has a non-vanishing $\df_\perp^a$. We can write the norm of the turning rate, $\omega \equiv \left| \omega^a \right|$, as:
\begin{align}
\frac{\omega}{H} = -\frac{\hat{s}^a V_a}{H \dot\phi} = \frac{\dot\phi_\perp V_v}{H \dot\phi^2}.
\label{eq:omegaDefn}
\end{align}
\subsection{Re-expressing the turning rate}\label{sec:TurningRate}
Ideally, we seek an expression for $\omega/H$ in terms of only the potential and its derivatives, which would allow us to bound when slow-roll, rapid-turn inflation is possible without evolving the equations of motion.
In other words, we would like to eliminate all velocity dependence from \eqref{eq:omegaDefn}.
In this subsection, we derive several expressions which nearly achieve this goal: the velocity dependence is reduced to only the direction of the perpendicular velocity $\perp^a \equiv \df_\perp^a / \df_\perp$.
In the restriction to two fields, this velocity dependence vanishes (as there is only one direction perpendicular to the potential gradient), and we recover Bjorkmo's results.

We begin by examining the $\omega$-dependence of the norm of the velocities. Using the equations of motion, we may express them exactly in terms of $\epsilon$, $\eta$, and $\omega$:
\begin{equation}
\begin{aligned}
\frac{\dot\phi_v}{\dot\phi} &= \frac{-3\, ( 1- \epsilon/3 + \eta/6)}{\sqrt{9 ( 1- \epsilon/3 + \eta/6)^2 + \omega^2/H^2}} \\
\frac{\dot\phi_\perp}{\dot\phi} &= \frac{\omega/H}{\sqrt{9 ( 1- \epsilon/3 + \eta/6)^2 + \omega^2/H^2}} \\
\dot\phi &= \frac{V_v}{H \sqrt{9 ( 1- \epsilon/3 + \eta/6)^2 + \omega^2/H^2}}.
\end{aligned}
\label{eq:vperp_eom}
\end{equation}
Slow-roll, rapid-turn trajectories  have both $\epsilon \ll 1$ and $|\eta| \sim \cO(\epsilon) \ll 1$, so we may drop the explicit $\epsilon$ and $\eta$ terms in \eqref{eq:vperp_eom} for velocities accurate to $\cO(\epsilon)$, i.e. $\df_v = \dbf_v+\cO(\epsilon)$.
When this truncation is consistent, the accelerations terms are negligible: $\partial_t \dbf \sim \cO(\epsilon H \dbf )$. Alternatively, the $v^a$ and $\perp^a$ acceleration components,
\begin{align}
    \frac{\partial_t \dbf_v}{H \dbf_v}  &= \frac{V_{vv} \df_v + V_{v\perp} \df_\perp}{H V_v} + \epsilon - \frac{2\nu}{\frac{9H^2}{\omega^2} + 1} \\
    \frac{\partial_t \dbf_\perp}{H \dbf_\perp}  &= \nu + \frac{\partial_t \df_v}{H \df_v}
\end{align}
must both be $\cO(\epsilon)$. Here, $\nu$ is given by:
\begin{align}
\nu &\equiv  \(\frac{\omega}{H} \)^\prime \bigg/ \(\frac{\omega}{H} \) \label{eq:nuDefinition} \\
&=\frac{V_{vv} \df_v + V_{v\perp} \df_\perp}{H V_v} + \cO(\epsilon),
\end{align}
where primes denote the e-fold derivative $\partial_N \equiv H^{-1}\partial_t$. A slowly-varying turn rate is therefore necessary for our truncated velocities to be consistent.


Combining these velocities and the equation of motion \eqref{eq:GradEOM}, we derive several expressions for $\omega$:
\begin{align}
\frac{\omega}{H} &= -\frac{3\dot{\phi}_\perp}{\dot{\phi}_v} \(1-\frac{\epsilon}{3}+\frac{\eta}{6} \), \label{eq:omegaVelocitiesExact} \\
\frac{\omega}{H} &= \frac{\perp^a V_{;ab} \dot\phi^b}{H V_v} + \mathcal{O}(\epsilon), \label{eq:omega2} \\
\frac{\omega^2}{H^2} &= \frac{V_{\perp \perp}}{H^2} - \frac{3}{\omega}\frac{V_{v\perp}}{H} - 9 + \mathcal{O}(\epsilon) \label{eq:omegaHessianGeneral}
\end{align}
where $\perp^a \equiv \dot\phi_\perp^a / \dot\phi_\perp$ and $V_{\perp b} \equiv \perp^a V_{;ab}$.
Further, when we demand the turning rate to be slowly-varying, i.e. $\nu \sim \cO(\epsilon)$, the last expression reduces to two explicit equations:
\begin{align}
	\frac{\omega}{H} &= \frac{3 V_{vv}}{V_{v\perp}} + \mathcal{O}(\epsilon) \label{eq:omegaHessianSimple} \\
	\frac{\omega^2}{H^2} &= \frac{V_{\perp \perp}}{H^2} - \frac{V_{v\perp}^2}{V_{vv} H^2} - 9 + \mathcal{O}(\epsilon). \label{eq:Bjorkmo13}
\end{align}
These must agree in regions of the potential where a long-lasting rapid-turn solution is possible.

The difference between two dimensions and higher dimensions is manifest in (\ref{eq:omega2})-(\ref{eq:Bjorkmo13}). At first glance, these expressions allow us to compute $\omega$ purely from components of the covariant Hessian matrix $V_{;ab}$, ostensibly satisfying our goal of identifying regions of field space admitting rapid-turn solutions. However, with more than two fields, the unit vector $\perp^a$ is not uniquely determined by $v^a$: it requires the velocity direction as well. Fortunately, we find that the velocities $\df_\perp^a$ may be computed algebraically from the equations of motion upon imposing carefully formulated assumptions of slow-roll and sufficiently high turning\footnote{Some limited information about these Hessian components is available without knowing $\perp^a$. Because the Hessian is a linear transformation, $\lambda_\mathrm{min} \leq  V_{;\perp\perp} \leq \lambda_\mathrm{max}$ and $V_{;v \perp }^2 \leq V_{;v a} V_{;v b} G^{ab}$, where $\lambda_\mathrm{min}$ and $\lambda_\mathrm{max}$ are the minimum and maximum eigenvalues of the Hessian. }.

\subsection{Extreme turning and the perpendicular velocity}

We now seek a slow-roll, rapid-turn solution for $\dot{\phi}_\perp^a$ by solving the equations of motion \eqref{eq:PerpEOM} and \eqref{eq:GradEOM}.
We first consider the acceleration terms and notice that not all components of $\ddot{\phi}_\perp^a$ can be neglected. Using \eqref{eq:omega2},  we can see that $ \ddot{\phi}_{\perp}^{a} \supset v^a \df_\perp \omega $.
Fortunately, this term does not enter the low-order slow-roll parameters.
The acceleration parameter $\eta$ can be expanded as:
\begin{equation}
    \begin{aligned}\label{eq:etaSlowRoll}
    \eta &= \frac{\cD_t(\dot\phi^2)}{2 \epsilon H^3}+2\epsilon \\
    &= -6 - \frac{V_v \dot{\phi_v}}{\epsilon H^3 } + 2 \epsilon, \\
    \end{aligned}
\end{equation}
where
\begin{equation}
\begin{aligned}\label{eq:DtPhiDotSquared}
    \cD_t(\dot\phi^2)
    = 2 (\dot\phi_v \ddot\phi_v + \dot\phi_{\perp a} \ddot\phi_\perp^a).
    \end{aligned}
\end{equation}
In terms of the accelerations, demanding $\eta\sim \cO(\epsilon)$ without a fine-tuned cancellation then requires:
\begin{align}
    \frac{\dot{\phi}_v \ddot{\phi}_v}{H^3} \sim \frac{\dot\phi_{\perp a} \ddot\phi_\perp^a}{H^3} \lesssim \cO(\epsilon^2).
\end{align}
Notably, the gradient acceleration $\ddot\phi_v$ and the component of $\ddot\phi_\perp^a$ along $\perp^a$ are constrained by $\eta$. Observing that trajectories with nontrivial turning may have an appreciable $\df_v$ via \eqref{eq:omegaVelocitiesExact}, we are motivated to neglect the acceleration term $\ddot\phi_v$ for full generality. Doing so renders \eqref{eq:GradEOM} a purely algebraic equation for the velocity $\df_v$.

Unfortunately, this is not sufficient to solve the other equation of motion \eqref{eq:PerpEOM}.
The accelerations in directions aligned with neither the perpendicular velocity nor the potential gradient could in principle be arbitrarily large, yet not contribute to $\eta$.
These accelerations cannot be trivially neglected in a field space of dimension $N_f > 2$; this contrasts the two-field attractor discussed in \cite{Bjorkmo:2019fls}, where these accelerations are trivially absent. Their magnitudes are determined by the potential's covariant Hessian, which we now proceed to analyze more closely.

We parametrize these directions with the projector $\Pi^{ab} \equiv \perp^{ab} - \perp^a \perp^b$ and define an orthonormal basis $\{\Pi^a_{(p)} |p=1,\ldots,N_f-2 \}$ spanning this $\Pi$-subspace. These directions have a contribution to the equations of motion given by:
\begin{align}
\ddot\phi_{\perp(p)} \equiv \Pi_{(p)a} \ddot\phi_\perp^a = -\frac{V_{;(p)c} \df^c}{V_v} \df_v ,
\label{eq:extraPerpAccel}
\end{align}
where $V_{;(p)c} \equiv \Pi^b_{(p)} V_{;bc} $.
At least a priori, we cannot rule out slow-roll, rapid-turn initial conditions with these terms arbitrarily large.

Nevertheless, we know that the gradient component of the acceleration $\ddot\phi_\perp^a$ is large during rapid-turn trajectories: $\ddot\phi_\perp^a v_a/(H \df_\perp) = \omega/H + \cO(\epsilon)$.
By comparison, the acceleration components in the $\Pi$-subspace are inversely proportional to $\omega$:
\begin{align}
\frac{\ddot\phi_{\perp(p)} }{H \df_\perp} &= \frac{-9H V_{;v(p)} + 3\omega V_{;\perp (p)} }{9H^2 \omega + \omega^3} + \cO(\epsilon) .
\label{eq:extraPerpAccelOmega}
\end{align}
We observe that \eqref{eq:omegaHessianGeneral} ensures $\omega$ is independent of the Hessian's contractions in the $\Pi$-subspace, $V_{;(p)a}$. We therefore expect these $\Pi$-accelerations to be suppressed in the limit of an extremely large $\omega$. This limit holds so long as the Hessian components $V_{;(p)a}/H$ are sufficiently small with respect to $\omega$.

In order to constrain these Hessian components, we define $\ddot\phi_{\perp v} \equiv v_a\ddot\phi^a_\perp $ and require the norm-square of $\ddot\phi_{\perp (p)} $ be smaller than that of $\ddot\phi_{\perp v}$. Demanding that $\nu$ be small, we replace the hessian components in terms of $\omega$ using \eqref{eq:omegaHessianSimple}-\eqref{eq:Bjorkmo13}. Then, expanding in powers of $\omega$, we have:
\begin{align}
	\frac{\ddot\phi_{\perp \Pi}^2 }{\ddot\phi_{\perp v}^2} = \frac{9H^2 V_{;\perp \Pi}^2 }{\omega^6} - \frac{54H^3 V_{;\perp\Pi}V_{;\Pi v} }{\omega^7} + \frac{81H^4 V_{;v\Pi}^2 - 162 H^4 V_{;\perp\Pi}^2 }{\omega^8} + \cO\(\frac{1}{\omega^9} \) + \cO\(\epsilon \),
\end{align}
where $\ddot\phi_{\perp\Pi}^2 \equiv \ddot\phi_\perp^a \Pi_{ab} \ddot\phi_\perp^b$ and $V_{;a \Pi}V_{;\Pi b} \equiv V_{;a c} \Pi^{cd} V_{;db}$. Barring a fine-tuned cancellation, the $\Pi$ components are negligible compared to the gradient component when:
\begin{align}\label{eq:HessianPiConstraints}
	\frac{V_{;\perp \Pi}^2 }{H^4} \ll \frac{\omega^6}{H^6},
	\quad \frac{V_{;v \Pi}^2 }{H^4} \ll \frac{\omega^8}{H^8}.
\end{align}
On the other hand, the analogous ratio involving $\ddot\phi_{\perp\perp} \equiv \perp_a\ddot\phi^a_\perp $ vanishes to $\cO(\epsilon)$ so long as $\nu$ is sufficiently small to admit \eqref{eq:omegaHessianSimple}:
\begin{align}
	\frac{\ddot\phi_{\perp\perp} }{\ddot\phi_{\perp v}} \simeq \cO\(\epsilon \).
\end{align}
This highlights the subtle, yet critical, connection between the longevity of turning $\nu$ and the longevity of inflation $\eta$ we first observed in Section \ref{sec:TurningRate}.

These bounds illustrate the conditions required for a high-turning trajectory to have negligible $\Pi$-accelerations. In this limit, the only non-negligible acceleration is $\ddot\phi_{\perp v}$. All other projections of the equations of motion \eqref{eq:PerpEOM} and \eqref{eq:GradEOM} may therefore be treated as algebraic equations for the velocities $\df_\perp^a$ and $\df_v$. We dub this the \textit{extreme turning} limit.

Since the Hessian contractions in \eqref{eq:HessianPiConstraints} involve the $\perp^a$ direction, we cannot immediately determine what qualifies as extreme turning from the potential and metric alone. We can, however, define the threshold value,
\begin{align}
	\frac{\omega_\mathrm{extreme} }{H} \equiv \( \frac{V_{;va}\perp^{ab}V_{;bv} }{H^4} \)^{1/8},
\end{align}
which is strictly larger than the Hessian contraction $( V_{;v\Pi}^2/H^4 )^{1/8}$ appearing in \eqref{eq:HessianPiConstraints}. The gradient direction and the projector $\perp^{ab}$ are determined exclusively from the potential and metric, so $\omega_\mathrm{extreme}$ can be computed without having a solution at hand. Extreme turning trajectories then satisfy $\omega \gg \omega_\mathrm{extreme}$.

The methods described below will only search the subset of all possible solutions that feature extreme turning. We treat \eqref{eq:HessianPiConstraints} as self-consistency conditions that may be verified after solving the equations of motion for the velocities $\df_\perp^a$ via our methods. In the only explicit $N_f \geq 3$ sustained rapid-turn trajectory we know of, all components of $\ddot{\phi}_{\perp(p)}$ are sufficiently small to admit $\omega_\mathrm{extreme} \ll \omega$;  see Section \ref{sec:helix}. However, we suspect that \eqref{eq:HessianPiConstraints} could be violated and still lead to a slow-roll, rapid-turn trajectory. Regrettably, we lack the tools to explore these general conditions. 

With this in mind, we restrict our search to initial conditions with the perpendicular accelerations $\ddot\phi^a_\perp$ approximately aligned with $v^a$. In other words, we take $\ddot{\phi}_\perp^a \dot{\phi}_{\perp a} \approx 0$ not by projection, but by zeroing all components of $\ddot{\phi}_\perp^a$ perpendicular to the gradient.

Under this assumption, the terms in \eqref{eq:PerpEOM} in the direction perpendicular to the gradient cancel:
\begin{align}\label{eq:slowRollPerpCancel}
	3H V_v \dot{\phi}_\perp^a \approx -\perp^{ab}V_{;bc}\dot{\phi}^c \dot{\phi}_v.
\end{align}
Expanding $\dot{\phi}^c = \dot{\phi}_v v^c + \dot{\phi}_\perp^c$, we find:
\begin{align} \label{eq:slowRollPerpEOM}
	\left( 3H V_v \delta^a_c + \dot{\phi}_v \perp^{ab} V_{;bc} \right) \dot{\phi}_\perp^c = -\dot{\phi}_v^2 \perp^{ab} V_{;bc} v^c,
\end{align}
which can be solved as a matrix equation for the components $\df_{\perp}^c$. In practice, a simultaneous method of finding $\df_v$ is also necessary -- we describe our numerical technique below in Section \ref{sec:numericalmethod}.

Solving \eqref{eq:slowRollPerpEOM} can give multiple solutions, which we now enumerate.
To solve for the velocities, we must invert the matrix on the l.h.s. of \eqref{eq:slowRollPerpEOM}, then plug the solution for $\df_\perp^a$ into the $\df_v$ equation of motion.
The inverse matrix will be inversely proportional to the determinant, which is a polynomial of degree $\df_v^{N_f-1}$. The entries in the matrix of cofactors will be polynomials of degree  $\df_v^{N_f-2}.$
Then, naively, the $\df_v$ equation (\ref{eq:GradEOM}) is an $\cO(\df_v^{2N_f})$ polynomal, and could have up to $2N_f$ real solutions.
However, the highest power coefficient is always zero, as can easily be seen by setting $H=0$ in (\ref{eq:slowRollPerpEOM}), and the polynomial is only of order $2N_f -1$.
Though $2N_f-1$ solutions are possible, rarely in practice are all of these real-valued. As we show in Section \ref{sec:alignedHessian}, the number of physically possible solutions is closely linked to the eigenvalue spectrum of the Hessian.

\subsection{Longevity of solutions}

All real-valued solutions correspond to valid rapid-turn initial conditions, though are not guaranteed to be long-lived trajectories -- i.e., there is no guarantee $V_{vv}$, $V_{\perp v}$, and $V_{\perp\perp}$ maintain a high $\omega$ at other points along the trajectory.
In our searches, we estimate the longevity of a solution by computing higher-order slow-roll parameters. We use $\nu$ as defined in \eqref{eq:nuDefinition} and
\begin{align}
\xi &\equiv\eta^\prime/\eta.
\label{eq:xiDefinition}
\end{align} When both $\nu$ and $\xi$ are small, those initial conditions will correspond to a trajectory that remains slow-roll and rapidly turning. Note that $\eta$ is fixed to be $\cO(\epsilon)$ by our method of solving for the velocities.
Using \eqref{eq:omegaDefn} and the equation of motion \eqref{eq:PerpEOM}, we recover the expression for $\nu$ as shown by Bjorkmo \cite{Bjorkmo:2019fls}, but generalized for any number of fields:
\begin{align}\label{eq:omegaChangeExact}
\(\frac{\omega}{H} \)' = -\mu_\times + \frac{\omega}{H}\(-3+\epsilon-\eta \),
\end{align}
where
\begin{align}\label{eq:muCross}
\mu_\times = \frac{\df_{\perp} \df_v}{H^2\df^2} \( V_{\perp\perp} - V_{vv} \) + \frac{V_{\perp v}}{H^2\df^2} \(\df_v^2 - \df_\perp^2 \) - 2\frac{\omega}{H}\epsilon.
\end{align}
Alternatively $\nu$ can be expressed as
\begin{align}
\nu = \frac{\ddot{\phi}_\perp^a \df_{\perp a}}{H \df_\perp^2} + \frac{\df_\perp V_{\perp v}+ \df_v V_{vv}}{H V_v} + 3 \epsilon - \eta.
\label{eq:otherNu}
\end{align}
Similarly,
\begin{equation}
\begin{aligned}
\xi &=\frac{\ddot{\phi}_v V_v}{H^4 \epsilon \eta} + \df_v\frac{\df_\perp V_{\perp v}+ \df_v V_{vv}}{H^4 \epsilon \eta} + \eta  -3\epsilon - 18\epsilon/\eta + 6\epsilon^2/\eta + 6 \\
&= \frac{\ddot\phi_v^2 + \ddot\phi_\perp^a \ddot\phi_{\perp a} + \df_v \dddot{\phi}_v + \df_\perp^a \dddot{\phi}_{\perp a} }{\eta \epsilon H^4} + 12\epsilon - 12\frac{\epsilon^2}{\eta} - 2 \eta
\end{aligned}
\end{equation}
where the two expressions for $\xi$ come from differentiating the two expressions for $\eta$ in \eqref{eq:etaSlowRoll}.
These parameters can be expanded to leading order in $\epsilon$
\begin{align}
\nu &= \frac{V_{vv} \dot\phi_v + V_{v\perp} \dot\phi_\perp}{H V_v} + \cO(\epsilon)\\
\xi &=  \nu \frac{V_v\df_v}{H^3} \times \cO(\epsilon^{-2}) = -6\nu \times \cO(\epsilon^{-1})
\end{align}
which makes it apparent that $\xi$ cannot be estimated to $\cO(1)$ with velocities less accurate than $\cO(\epsilon^2)$. Nonetheless, we have found $\xi$ to be a helpful discriminator between solutions.

Although $V_{vv}$ is absent from the equations of motion, we see its relevance here: long lived solutions will have a $V_{vv}$ and $V_{v\perp}$ such that the leading order term in $\nu$ approximately cancels.
In fact, one particularly effective check that a solution is long-lasting is to compare the various expressions for $\omega^2$ in Section \ref{sec:TurningRate} -- when $\nu$ is small and all of the accelerations are negligible, they will all agree to $\cO(\omega \epsilon)$.

\subsection{Numerical method}
\label{sec:numericalmethod}
To numerically determine where a model satisfies rapid-turn inflation, we perform the following steps:
\begin{itemize}
\item After inputting the potential and metric, perform the necessary symbolic manipulations to compute $v^a$, $V_v$, and $V_{;ab}$.
\item Choose a point in field space, and explicit values for any parameters in the potential.
\item Guessing $\dot\phi_v$,  multiply \eqref{eq:slowRollPerpEOM} by the inverse of the matrix on the l.h.s. to solve for $\dot{\phi}_\perp^a$. Using \eqref{eq:GradEOM}, check the acceleration $\ddot{\phi}_v$ implied by this value of $\dot\phi_v$.
\item Repeat the last step, choosing $\dot\phi_v$ in the range $0 < -\dot{\phi}_v < H\sqrt{2 \epsilon_V}$ to give $\ddot{\phi}_v = 0$.
In practice, we find the zeros of $\ddot{\phi}_v / \dot{\phi}_\perp^2$ and use a bisection algorithm to identify them to high precision. There will always be exactly $2N_f -1$ values of $\dot{\phi}_v$ that set $\ddot{\phi}_v$ to zero, though none of them are guaranteed to lie within the searched range.
\item Using the velocities corresponding to all found zeros of $\ddot{\phi}_v$, compute the various slow-roll parameters $\epsilon,\omega,\eta,\nu,\xi,\ldots$ and compute and check the consistency conditions \eqref{eq:HessianPiConstraints}.
\item Repeat the previous four steps, scanning the model's field and parameter space. Points that support extended rapid-turn solutions are assumed to minimize a cost function
\begin{align}
\mathrm{cost} \equiv \left| \frac{\ddot{\phi}_v}{3 H \dot{\phi}_v} \right| + A \epsilon + B (|\eta_1| + |\eta_2|) + C |\nu| +\frac{D}{\omega^2} + \frac{E}{\epsilon_V}
\label{eq:cost}
\end{align}
where the two expressions for $\eta$ are those in \eqref{eq:etaSlowRoll} and $A,B,\ldots,E$ are arbitrary relative weights of each term. In our type-IIA constructions, we typically take $A = 1, B = C=0.1, D=E=0$ while in potentials that allow for slow-roll, slow-turn inflation, we take $D,E > 0 $.
We use a differential evolution optimizer to efficiently perform this scan over field space and parameter space \cite{Feldt2018}.
\end{itemize}
Our implementation of this strategy was created using one of the authors's Julia-language inflation code, \texttt{Inflation.jl}\footnote{\url{https://github.com/rjrosati/Inflation.jl}}.
Despite the complexity of finding rapid-turn solutions this way, it remains orders of magnitude faster than blindly evolving the equations of motion.

\subsection{Attractor behavior}
In order to analyze the attractor nature of multi-field inflationary trajectories, we require the mass matrix of a trajectory's perturbations. The full mass matrix reduces to block-diagonal form when certain perturbative modes decouple from the remaining degrees of freedom, thus drastically simplifying the analysis. This is the case in generalized hyperinflation \cite{Bjorkmo:2019aev}, where perturbations along the adiabatic and first entropic directions decouple from the higher entropic modes. Subsequently, only the $2\times 2$ block of the full mass matrix describing these two modes need be considered to determine stability of the trajectory. This decoupling is also manifest in the time derivatives of the gradient direction $v^a$ and a single perpendicular direction $w^a$, where one exclusively sources the other.

Assuming an arbitrary potential and field space geometry, we do not find a simple decoupling of two modes from the remaining degrees of freedom. The gradient's time derivative is given by
\begin{align}\label{eq:gradientDerivative}
\cD_t v^a = \perp^{ab}\frac{V_{;bc}\df^c}{V_{;v}} = \perp^{ab}\frac{V_{bc}}{V_{;v}}\(\df_v v^c + \df_\perp\perp^c \)
\end{align}
and mixes the gradient with the (a priori) unknown perpendicular direction $\perp^a$. Due to the generic structure of the Hessian matrix, we cannot readily identify an a priori known direction along which this time derivative lies. This contrasts generalized hyperinflation, where $\cD_t v^a$ can be made proportional to a particular eigenvector of the Hessian, $w^a$, and the remaining eigenvectors' derivatives fully decouple from these two vectors. We observe that this is possible in the extreme turning limit, since the $\Pi$-subspace components of the gradient's derivative are negligible. Then, the right hand side of \eqref{eq:gradientDerivative} reduces to $\omega \perp^a$. We leave an analysis of this specific limit's perturbations and attractor behavior for future work.

As a result, the perturbations' equations of motion do not feature generalized hyperinflation's decoupling of adiabatic and first entropic modes from the higher entropic modes. Hence, the fully coupled equations do not readily admit an analytically simple mass matrix. In light of this, we believe the attractor nature of a high-turning trajectory for a generic potential and geometry must therefore be verified numerically.

Due to this absence of decoupling, we do not have a general prescription for analyzing the phenomenology of perturbations in any potential and geometry. However, we do observe that trajectories with high rates of turning are known to feature a tachyonic instability that leads to an exponential growth of subhorizon modes' power spectra \cite{Fumagalli:2019noh,Bjorkmo:2019qno,Renaux-Petel:2015mga,Garcia-Saenz:2018ifx,Garcia-Saenz:2018vqf}. This growth must be considered in any phenomenological study of rapid-turn trajectories. We end by noting that phenomenologically viable rapid-turn trajectories have been found in \cite{Aragam:2019khr,Garcia-Saenz:2018ifx,Garcia-Saenz:2018vqf}, in which the mass scale of inflation is lowered to ensure an adiabatic power spectrum that is consistent with observations.


\section{The Aligned Hessian Approximation}
\label{sec:alignedHessian}

The equations of motion \eqref{eq:PerpEOM} and \eqref{eq:GradEOM}  yield particularly simple analytic results when the gradient is an eigenvector of the potential's covariant Hessian matrix. As we argue below in Section \ref{sec:whenHessAlign}, this is generic in hyperbolic  two-dimensional field space. The resulting velocities can subsequently be used to compute the turning rate analytically. As an example, hyperinflation and its generalizations \cite{Brown:2017osf,Mizuno:2019pcm,Bjorkmo:2019aev} (also, Section \ref{sec:hyperinflation}) are a well-studied class of high-turn rate models with such a Hessian.

For future use, we define an orthonormal basis $\{w_{(i)}^a|i=1,\ldots,N_f-1\}$ that spans the directions in field space orthogonal to the gradient such that $\perp^{ab} = \sum_{i} w_{(i)}^a w_{(i)}^b$.
For now, we make no assumption about the alignment of this basis with respect to the Hessian eigenvectors. In this basis, \eqref{eq:PerpEOM} and \eqref{eq:GradEOM} reduce to:
\begin{align}
	\ddot\phi_{\perp i} &= -\left[ 3 H \delta_{ij} + \frac{\dot\phi_v}{V_v} V_{;ij} \right] \dot\phi_{\perp j} - \frac{\dot{\phi}_v^2}{V_v} V_{;iv} \label{eq:PerpProjectedEOM} \\
	\ddot\phi_v &= -\left[ 3 H - \frac{\df_\perp^i V_{;iv}}{V_v} \right] \dot\phi_v - V_v + \frac{\df_\perp^i V_{;ij} \df_\perp^j}{V_v}. \label{eq:GradProjectedEOM}
\end{align}

\subsection{When does the Hessian align?}
\label{sec:whenHessAlign}
We have observed that in large regions of parameter space, the gradient of the potential aligns closely with one of the Hessian eigenvectors. We lack a formal proof of this alignment for an arbitrary potential and geometry, but our numerical scans of parameter space indicate this holds across many different models. In Section \ref{sec:examples}, we present several examples in which this alignment yields robust results.

There is, however, a logical explanation for this alignment in the case of a two-dimensional hyperbolic metric. Let

\begin{equation} G_{ab} =
\left(
\begin{array}{cc}
 1&0 \\
0 & \sinh^2 x
\end{array}
\right) 
\end{equation}

In the limit $ x \gg1 $, a vector $V=(v_1,v_2)$ when normalized takes the form $\hat{V} = (1, \frac{v_2}{v_1}) + O\left(\frac{1}{\sinh^2 x }\frac{v_2^2}{v_1^2}\right)$. The scalar product of two normalized vectors: $\hat{V} $ and $\hat{U}$, in the region of $x \gg 1$ is approximately:

\begin{equation}
\hat{V} \cdot \hat{U} = 1 + O\left(\frac{1}{\sinh^2 x} \frac{v_2 \, u_2}{v_1 \, u_1}\right)
\end{equation}
There is alignment as long as 

\begin{equation}
\left(\frac{1}{\sinh^2 x} \frac{v_2 \, u_2}{v_1 \, u_1} \right) \ll 1
\end{equation}
This relation will not hold for the normalized orthogonal vector to $\hat{V} $ as

\begin{equation}
\hat{V}_{\perp} =\left( \frac{1}{\sinh x} \, \frac{v_2}{v_1}, - \sinh x  \right)
\end{equation}

Conversely, in the limit $ x \ll 1 $ a vector $V=(v_1,v_2)$ when normalized takes the form $\hat{V} = x ( \frac{v_1}{v_2},1) + O\left(x^2 \right)$. The scalar product of two normalized vectors: $\hat{V} $ and $\hat{U}$, in the region $x \ll 1$ can be approximated by. 

\begin{equation}
\hat{V} \cdot \hat{U} = 1 + O\left( x^2 \frac{v_1^2}{v_2^2}\ \right) \sim 1
\end{equation}
Thus, treating the gradient as an eigenvector of the Hessian should be a good approximation everywhere but the region around $x\sim 1$. When the metric is of the form,

\begin{equation} G_{ab} =
\left(
\begin{array}{cc}
 e^{2 y/ R_0}&0 \\
0 &1
\end{array}
\right),
\end{equation}
similar reasoning to the one given above explains why the gradient is close to being an eigenvector of the Hessian when $y \gg R_0$. 

In the higher dimensional spaces that we study in sections 4.3-4.8, there are also sizable regions of field space where the gradient is parallel to an eigenvector of the Hessian. A careful analysis shows that this result depends both on the metric and the potential.

%

\subsection{The diagonal Hessian}
For this subsection, we'll assume the Hessian has an eigenvector exactly aligned with $v^a$. Then the $w^a_{(i)}$ align with the Hessian eigenvectors, and the Hessian is diagonal in the $\{v,w_{(i)}\}$ basis:
\begin{align}\label{eq:gradDiagHess}
V^{;ab} = V_{;vv} v^a v^b + V_{;w_1 w_1} w_1^a w_1^b + V_{;w_2 w_2} w_2^a w_2^b + \ldots
\end{align}
Here, the Hessian's eigenvalues are not required to be equal in the perpendicular directions.

Under this assumption, $V_{;iv} = 0$ and $V_{;ij} = \lambda_i \delta_{ij}$, with no summation implied. The equations of motion \eqref{eq:PerpProjectedEOM} and \eqref{eq:GradProjectedEOM} become
\begin{align}
\ddot{\phi}_v &= -3 H \dot\phi_v - V_v + \frac{\dot\phi_\perp^i V_{;ij} \dot\phi_\perp^j}{V_v} \\
\ddot{\phi}_{\perp i} &= -\left[ 3H  + \frac{\dot\phi_v}{V_v} \lambda_i \right] \dot\phi_{\perp i}.
\label{eq:diagHessPhipEOM}
\end{align}

Note that the $\Pi$ accelerations defined earlier (c.f. \eqref{eq:extraPerpAccel}) are identically zero with this Hessian structure.

When $\eta$ is small and the accelerations are negligible, two types of solution are possible. Either all components of $\dot\phi_{\perp i}$ are zero and we recover slow-roll ($3 H \dot\phi_v = - V_v $), or, for each component of $\dot\phi_{\perp i}$, either $\dot\phi_v = -3 H V_v / \lambda_i$ or $\dot\phi_{\perp i} = 0$. Since $\dot\phi_v$ is a scalar, only one eigenvalue, $\lambda_k$, can contribute to its value:
\begin{align}\label{eq:gradVelGradEvec}
	\df_v = -\frac{3HV_v}{\lambda_k}.
\end{align}
Multiple components of $\dot\phi_{\perp i}$ could be nonzero if some of the Hessian's eigenvalues were degenerate, but all other components of $\dot\phi_{\perp i}$ in directions with different eigenvalues $\lambda_i \neq \lambda_k$ must vanish. We denote $d$ as the degeneracy of $\lambda_k$.

For any amount of degeneracy $d$, we can solve the $\dot\phi_v$ equation to get
\begin{align}
\dot\phi_{\perp}^2 &= \left(1 - \frac{9 H^2}{\lambda_k} \right) \frac{V_v^2}{\lambda_k}, \label{eq:perpVelDegenerate}\\
\df^2 &= \df_v^2 + \df_\perp^2 = \frac{V_v^2}{\lambda_k},\label{eq:velSquaredGradEvec}
\end{align}
where $\dot\phi_{\perp}^2 = \dot\phi_{\perp k}^2$ when $d=1$, or a sum of the components with identical eigenvalues when not \footnote{Interestingly $\dot\phi_\perp$ has a maximum when $\lambda_k=18 H^2$.}. In either case, we get that 
\begin{align}
\epsilon &= \frac{ \dot\phi_v^2 + \dot\phi_\perp^2 }{2 H^2} = \frac{V_v^2}{2 H^2 \lambda_k} = \frac{3 V_v^2}{2 V \lambda_k+V_v^2} \sim \frac{3 V_v^2}{2 V \lambda_k} = \epsilon_V \, \frac{ 9 H^2}{\lambda_k}.
\label{eq:ehEigenvalue}
\end{align}
Comparing to \eqref{eq:omegaDefn}, we see that
\begin{align}\label{eq:omegaEigenvalue}
\frac{\omega^2}{H^2} = \frac{\lambda_k}{H^2} - 9
\end{align}
For these results to make sense, $\lambda_k$ must be positive and greater than $9 H^2$.
At a given point, all sufficiently large eigenvalues except for $\lambda_u$ could correspond to $\lambda_k$.
Together with the slow-roll solution and the two possible signs of $\df_{\perp k}$, we reproduce the same counting as the full equations of motion: up to $2N_f-1$ distinct solutions are possible.
As in the generic hessian case, all real-valued solutions correspond to valid initial conditions, and the longevity of a particular set of initial conditions is measurable through $\nu$, $\xi$, and other higher-order slow-roll parameters.

We note that expressions \eqref{eq:gradVelGradEvec}, \eqref{eq:velSquaredGradEvec}, and \eqref{eq:omegaEigenvalue} have two-field counterparts \cite{Bjorkmo:2019fls}, which arose from enforcing the constraints $V_{vv}/H^2 \lesssim \cO(\epsilon\omega^2/H^2)$ and $V_{v\perp}/H^2 \lesssim \cO(\epsilon\omega/H)$ on the two-field version of \eqref{eq:Bjorkmo13}.
Since the diagonal Hessian has $V_{v\perp} = 0$, the latter bound is automatically satisfied; the former bound is satisfied for a sufficiently large $\lambda_k$. In light of this, the diagonal Hessian may be viewed as a subset of the two-field rapid-turn solution that generalizes to an arbitrary number of fields.

Neglecting the accelerations gives a consistent solution when the derivatives of the velocities are small $\partial_t \df \sim \cO(\epsilon H \df )$. We compute them as
\begin{align}
\label{eq:diag_accel_check}
\frac{\partial_t \df_v}{H \df_v} &= V_v^\prime/V_v + \epsilon - \lambda_k^\prime/\lambda_k \\
\frac{\partial_t \df_\perp}{H \df_\perp} &= \nu + \frac{\partial_t \df_v}{H \df_v} \\
\end{align}

They will be long-lived when the higher slow-roll parameters are small.
\begin{align}
\nu &= \frac{V_{vv} \df_v}{H V_v} + \frac{V_{v\perp}\df_\perp}{HV_v} + \cO(\epsilon) \\
&= - 3\frac{V_{vv}}{\lambda_k} + \frac{\omega}{H}\frac{V_{v\perp}}{\lambda_k} + \cO(\epsilon)\\
\eta &= -6\frac{V_{vv}}{\lambda_k} + 2 \frac{V_{v\perp} \df_\perp}{V_v H} + 3 \frac{\lambda_{k v} V_v}{\lambda_k^2} - \frac{\lambda_{k \perp} \df_\perp}{H \lambda_k} + 2\epsilon \label{eq:diagEtaGeneral} \\
&\simeq 2 \nu - \lambda_k^\prime / \lambda_k  + \cO(\epsilon) \\
&= 2 \nu  + \frac{3\lambda_{k v} V_v}{\lambda_k^2} - \frac{\lambda_{k \perp} V_v}{\lambda_k^2} \frac{\omega}{H} + \cO(\epsilon)
\label{eq:diag_eta_check}
\end{align}
where we computed $\eta$ directly from \eqref{eq:gradVelGradEvec}-\eqref{eq:velSquaredGradEvec} and have left the $V_{v \perp}$ terms in, and $\lambda_{kv} \equiv V_{;kkv}$, $\lambda_{k\perp} \equiv V_{;kk\perp}$. We have substituted in the velocity solutions directly.

From \eqref{eq:diag_accel_check}-\eqref{eq:diag_eta_check}, we see that $\cO(\epsilon)$-consistent solutions to the equations of motion will also necessarily have a small $\eta$: $\nu$ and $\lambda_k^\prime/\lambda_k$ set both. The lowest-order descriminator between solutions, then, is $\xi$.

\subsection{Non-diagonal corrections}
In light of \eqref{eq:omegaEigenvalue}, we seek to understand how \eqref{eq:gradDiagHess} changes when the gradient is strongly, but not exactly, aligned with an eigenvector of $V_{;ab}$. This will allow us to perturbatively expand this eigenvector about the gradient. Suppose $V_{;ab}$ has an eigenbasis $\{u^a,t_{(n)}^a|n=1,\ldots,N_f-1\}$ such that $u^a$ is approximately aligned with the gradient: $u^a v_a = 1-\alpha$, where $\alpha\ll 1$. We expand this eigenvector as:
\begin{align}\label{eq:eigvecHess1}
u^a = (1-\alpha)v^a + \sum_{i=1}^{N_f-1}\delta_i w_{(i)}^a.
\end{align}
Demanding $u^a u_a = 1$ and neglecting $\mathcal{O}(\alpha^2)$ terms fixes the sum-squared of coefficients:
\begin{align}\label{eq:eigvec1NormConstraint}
\sum_i \delta_i\delta_i \simeq 2\alpha.
\end{align}
Hence, each coefficient is bounded in magnitude: $\left|\delta_i\right| \lesssim \mathcal{O}(\alpha^{1/2})$.

Similarly, we expand the remaining eigenvectors: $t_{(n)}^a = \gamma_n v^a + \sum_i\beta_{ni} w_{(i)}^a$. Imposing $t_{(n)}^a u_a = 0$ fixes the $\gamma$ coefficients, $\gamma_n \simeq -(1+\alpha)\sum_i\delta_i\beta_{ni}$. This yields:
\begin{align}\label{eq:eigvecHessPerp}
t_{(n)}^a = \sum_{i=1}^{N_f-1}\beta_{ni}\[ -(1+\alpha)\delta_i v^a + w_{(i)}^a \] \equiv \sum_{i=1}^{N_f-1}\beta_{ni}q_{(i)}^a.
\end{align}
Note that the vectors $\{q_{(i)}^a\}$ are perturbatively orthogonal to $u^a$, i.e. $q_{(i)}^a u_a \simeq \mathcal{O}(\alpha^2)$. Hence, they span the same hyperplane as the vectors $\{t_{(n)}^a\}$. However, they are not mutually orthonormal to $\mathcal{O}(\alpha)$: $q_{(i)}^a q_{(j)a} \simeq \delta_{ij} + \delta_i\delta_j$. Therefore, we cannot view the coefficients $\beta_{ni}$ as enacting an orthogonal transformation. Orthonormality of $\{t_{(n)}^a\}$ yields the constraint:
\begin{align}
\sum_{i,j}\[ \beta_{ni}\beta_{mj}\delta_i\delta_j(1+2\alpha) + \delta_{ij}\beta_{ni}\beta_{mj} \] = \delta_{nm}.
\end{align}
Since $\delta_i\delta_j \sim \mathcal{O}(\alpha)$, we can neglect the combination $\delta_i\delta_j\alpha \sim \mathcal{O}(\alpha^2)$ to obtain:
\begin{align}\label{eq:orthonormConstraint1}
\sum_{i,j}\[ (\beta_{ni}\delta_i)(\beta_{mj}\delta_j) + \delta_{ij}\beta_{ni}\beta_{mj} \] = \delta_{nm}.
\end{align}
Henceforth, sums over repeated $\{i,j\}$ indices are implied except for evidently free indices. We can invert \eqref{eq:eigvecHess1} and \eqref{eq:eigvecHessPerp} to obtain expressions for the gradient and its perpendicular directions in the Hessian eigenbasis:
\begin{align}
v^a &= (1-\alpha)u^a - (1+\alpha)\sum_n^{N_f-1} (\beta_{ni}\delta_i)t_{(n)}^a\\
w_{(i)}^a &= \delta_i u^a + \sum_n^{N_f-1} \beta_{ni}t_{(n)}^a.
\end{align}
These vectors are orthonormal to $\mathcal{O}(\alpha)$ provided we enforce \eqref{eq:eigvec1NormConstraint} and orthonormality of the perpendicular directions: $w_{(i)}^a w_{(j)a} = \delta_{ij}$. This yields a second constraint:
\begin{align}\label{eq:orthonormConstraint2}
\sum_n\beta_{ni}\beta_{nj} = \delta_{ij} - \delta_i\delta_j.
\end{align}
The two orthonormality constraints \eqref{eq:orthonormConstraint1} and \eqref{eq:orthonormConstraint2} let us identify an explicit $\mathcal{O}(\alpha)$ expression for the matrices $\beta_{ni}$:
\begin{align}\label{eq:perpChangeOfBasis}
\beta_{ni} = \delta_{ni} - \frac{1}{2}\delta_n\delta_i.
\end{align}
Strictly speaking, the vectors $w_{(i)}^a$ need not be closely aligned with the eigenvectors $t_{(n)}^a$, since the $t_{(n)}^a$ need only be orthogonal to $u^a$ and slightly non-orthogonal to $v^a$. However, we are free to rotate the $w_{(i)}^a$ such that they are closely aligned with the $t_{(n)}^a$.

The Hessian is diagonal in the $\{u^a,t_{(n)}^a \}$ basis, so it takes the form:
\begin{align}
V_{;ab} = \lambda_u u_a u_b + \sum_{n=1}^{N_f-1}\lambda_n t_{(n)a}t_{(n)b},
\end{align}
where $\lambda_u$ and $\lambda_n$ are the eigenvalues of $u^a$ and $t_{(n)}^a$, respectively. Inserting \eqref{eq:eigvecHess1}, \eqref{eq:eigvecHessPerp}, \eqref{eq:perpChangeOfBasis}, and keeping only $\mathcal{O}(\alpha) \sim \mathcal{O}(\delta^2)$ terms, we find:
%
\begin{align}
\begin{aligned}\label{eq:hessDelta2}
V_{;ab} \simeq &\[ \lambda_u(1-2\alpha) + \sum_n \lambda_n\delta_n^2 \]v_a v_b\\
&+ \[ \lambda_u\delta_i\delta_j + \sum_n \lambda_n\( \delta_{ni}\delta_{nj} - \frac{\delta_i\delta_{nj}+\delta_j\delta_{ni}}{2}\delta_n \) \]w_{(i)a}w_{(j)b}\\
&+ \( \lambda_u - \sum_n \lambda_n\delta_n \)\delta_i \( v_a w_{(i)b} + w_{(i)a} v_b \).
\end{aligned}
\end{align}
This gives us the Hessian in a form that can be inserted into the equations of motion. Note that we are neglecting terms higher order in $\delta$ without considering the hierarchy of eigenvalues $\lambda_u$ and $\lambda_n$. We will later see that for $\lambda_u \gg \lambda_k$, this perturbative method becomes unreliable.

The components of $V_{;ab}$ entering \eqref{eq:PerpProjectedEOM} and \eqref{eq:GradProjectedEOM} simplify to:
\begin{align}
V_{;ij} &= \lambda_i\delta_{ij} + \( \lambda_u - \frac{\lambda_i+\lambda_j}{2} \)\delta_i\delta_j\\
V_{;iv} 
&= \( \lambda_u - \sum_n\lambda_n\delta_n \)\delta_i,
\end{align}
The equations of motion are then given by the $\mathcal{O}(\delta^2)$ expressions:
\begin{align}
\begin{aligned}\label{eq:perpExpandedEOM}
\ddot{\phi}_{\perp i} = -&\( 3H + \frac{\dot{\phi}_v}{V_v}\lambda_i \)\dot{\phi}_{\perp i}\\
&+ \frac{\dot{\phi}_v}{V_v}\sum_j\delta_i\( \lambda_u - \frac{\lambda_i+\lambda_j}{2} \)\delta_j \dot{\phi}_{\perp j} - \frac{\dot{\phi}_v^2}{V_v}\( \lambda_u - \sum_n\lambda_n\delta_n\ \)\delta_i
\end{aligned}
\end{align}
\begin{align}
\begin{aligned}\label{eq:gradExpandedEOM}
\ddot{\phi}_v 
= -&3H\dot{\phi}_v - V_v + \frac{1}{V_v}\sum_i\lambda_i\dot{\phi}_{\perp i}\dot{\phi}_{\perp i}\\
&+ \frac{1}{V_v}\sum_{i,j}\delta_i\dot{\phi}_{\perp i}\( \lambda_u - \frac{\lambda_i+\lambda_j}{2} \)\delta_j\dot{\phi}_{\perp j} + \frac{\dot{\phi}_v}{V_v}\( \lambda_u - \sum_n\lambda_n\delta_n \)\sum_i\delta_i\dot{\phi}_{\perp i}.
\end{aligned}
\end{align}
Setting the acceleration to zero in \eqref{eq:perpExpandedEOM} results in a matrix equation:
\begin{align}\label{eq:perpVelMatrixEqDelta2}
-\( 3H + \frac{\dot{\phi}_v}{V_v}\lambda_i \)\sum_j\[ \delta_{ij} - \frac{\dot{\phi}_v\(\lambda_u-\frac{\lambda_i+\lambda_j}{2} \)}{3HV_v+\lambda_i\dot{\phi}_v}\delta_i\delta_j \]\dot{\phi}_{\perp j} = \frac{\dot{\phi}_v^2}{V_v}\( \lambda_u - \sum_n\lambda_n\delta_n\ \)\delta_i.
\end{align}
The matrix multiplying $\dot{\phi}_{\perp j}$ is readily invertible, but the non-identity components yield $\mathcal{O}\(\delta^3 \)$ corrections to the right hand side. Dropping these, we arrive at an $\mathcal{O}\(\delta^2 \)$ expression for the perpendicular-direction velocities:
\begin{align}
\begin{aligned}\label{eq:perpExpandedEOMSoln}
-\( 3H + \frac{\dot{\phi}_v}{V_v}\lambda_i \)\dot{\phi}_{\perp i}
\simeq \frac{\dot{\phi}_v^2}{V_v}\(\lambda_u - \sum_n\lambda_n\delta_n \)\delta_i
\end{aligned}
\end{align}

We can solve \eqref{eq:perpExpandedEOMSoln} and \eqref{eq:gradExpandedEOM} with negligible accelerations by expanding the velocities in powers of $\alpha^{1/2}\sim\left|\delta_i \right|$:
\begin{align}
\df_{\perp i} &\simeq \df_{\perp i}^{(0)} + \alpha^{1/2}\df_{\perp i}^{(1)} + \alpha\df_{\perp i}^{(2)} + \cO\(\alpha^{3/2} \)\\
\df_{v} &\simeq \df_{v}^{(0)} + \alpha^{1/2}\df_{v}^{(1)} + \alpha\df_{v}^{(2)} + \cO\(\alpha^{3/2} \)
\end{align}
Inserting these into \eqref{eq:perpExpandedEOMSoln} leads to the order-by-order matching conditions:
\begin{align}
\hspace{6 pt}\(3H + \frac{\df_v^{(0)}}{V_v}\lambda_i \)\df_{\perp i}^{(0)} &= 0\label{eq:perpMatching0}\\
\alpha^{1/2}\[\(3H + \frac{\df_v^{(0)}}{V_v}\lambda_i \)\df_{\perp i}^{(1)} + \(\frac{\df_v^{(1)}}{V_v}\lambda_i \)\df_{\perp i}^{(0)} \] &= -\[\frac{\(\df_v^{(0)} \)^2}{V_v}\lambda_u \]\delta_i\label{eq:perpMatchingDelta}
\end{align}
\begin{align}
\begin{aligned}\label{eq:perpMatchingDelta2}
\alpha&\[\(3H + \frac{\df_v^{(0)}}{V_v}\lambda_i \)\df_{\perp i}^{(2)} + \(\frac{\df_v^{(1)}}{V_v}\lambda_i \)\df_{\perp i}^{(1)} + \(\frac{\df_v^{(2)}}{V_v}\lambda_i \)\df_{\perp i}^{(0)} \]\\
&= \frac{1}{V_v}\(\lambda_u - \frac{\lambda_i+\lambda_k}{2} \)\sum_j\df_v^{(0)}\df_{\perp j}^{(0)}\delta_i\delta_j - \[\frac{2\df_v^{(1)}\df_v^{(0)}}{V_v}\lambda_u \]\alpha^{1/2}\delta_i  + \[\frac{\(\df_v^{(0)} \)^2}{V_v}\sum_n\lambda_n\delta_n \]\delta_i
\end{aligned}
\end{align}
Likewise, \eqref{eq:gradExpandedEOM} gives the matching conditions:
\begin{align}
3H\df_v^{(0)} + V_v - \frac{\lambda_k}{V_v}\(\df_\perp^{(0)} \)^2 &= 0\label{eq:gradMatching0}\\
\alpha^{1/2}\[3H\df_v^{(1)} - \frac{2\lambda_k}{V_v}\sum_i\df_{\perp i}^{(1)}\df_{\perp i}^{(0)} \] &= \frac{\lambda_u}{V_v}\df_v^{(0)}\sum_i\df_{\perp i}^{(0)}\delta_i\label{eq:gradMatchingDelta}
\end{align}
\begin{align}
\begin{aligned}\label{eq:gradMatchingDelta2}
\alpha&\[3H\df_v^{(2)} - \frac{1}{V_v}\sum_i\(\lambda_i\df_{\perp i}^{(1)}\df_{\perp i}^{(1)} + 2\lambda_k\df_{\perp i}^{(2)}\df_{\perp i}^{(0)} \) \]\\
&= \frac{\lambda_u-\lambda_k}{V_v}\sum_{i,j}\df_{\perp i}^{(0)}\df_{\perp j}^{(0)}\delta_i\delta_j + \frac{\lambda_u}{V_v}\sum_i\(\df_v^{0}\df_{\perp i}^{(1)} + \df_v^{(1)}\df_{\perp i}^{(0)} \)\delta_i\alpha^{1/2} - \frac{1}{V_v}\sum_{n,i}\lambda_n\df_v^{(0)}\df_{\perp i}^{(0)}\delta_n\delta_i,
\end{aligned}
\end{align}
where $\lambda_k$ is the eigenvalue that solves $\df_v^{(0)} = -3HV_v/\lambda_k$, such that the corresponding velocity $\df_{\perp k}^{(0)} \neq 0$. For $\l \neq k$, we have $\df_{\perp l}^{(0)} = 0$, and the matching condition \eqref{eq:perpMatchingDelta} yields:
\begin{align}
\df_{\perp l}^{(1)} = -\frac{3HV_v\lambda_u}{\lambda_k\(\lambda_k-\lambda_l \)}\frac{\delta_l}{\alpha^{1/2}}.
\end{align}
For degeneracy $d$ of the eigenvalue $\lambda_k$, we see that \eqref{eq:perpMatching0} leaves the non-vanishing components $\df_{\perp k}^{(0)}$ unconstrained at zeroth order.  We can only solve for the norm-squared of the perpendicular velocity via \eqref{eq:gradMatching0}, as in \eqref{eq:perpVelDegenerate}. Assuming $d=1$, we can solve for the $k^{th}$ component:
\begin{align}
\df_{\perp k}^{(0)} = \pm\frac{V_v}{\lambda_k}\sqrt{\lambda_k-9H^2},
\end{align}
as well as the remaining $\cO\(\alpha^{1/2} \)$ terms:
\begin{align}
\df_v^{(1)} &= \mp\frac{\lambda_u}{\lambda_k^2}\frac{9H^2 V_v}{\sqrt{\lambda_k-9H^2}}\frac{\delta_k}{\alpha^{1/2}}\\
\df_{\perp k}^{(1)} &= \frac{3HV_v}{2}\frac{\lambda_u}{\lambda_k^2}\frac{\lambda_k-18H^2}{\lambda_k-9H^2} \frac{\delta_k}{\alpha^{1/2}}.
\end{align}
We therefore find the following $\cO\(\delta \)$ slow-roll expressions for $\epsilon$ and $\omega$:
\begin{align}
\label{eq:ehDelta}
\epsilon &= \frac{V_v^2}{2H^2\lambda_k} \pm \(\frac{\lambda_u}{\lambda_k^2\sqrt{\lambda_k - 9H^2}}\frac{3V_v^2}{2H} \)\delta_k.\\
\frac{\omega^2}{H^2} &= \frac{\lambda_k}{H^2} - 9 \mp \frac{\lambda_u}{9H\sqrt{\lambda_k-9H^2}}\delta_k.
\label{eq:omegaDelta}
\end{align}
The derivative of the turning rate \eqref{eq:omegaChangeExact} can also be computed to $\cO(\delta)$:
\begin{align}\label{eq:nuDelta}
\nu = \epsilon - 3\frac{\lambda_u}{\lambda_k} \pm \[\frac{\lambda_u}{\lambda_k}\frac{\lambda_k - 9H^2\frac{\lambda_u+\lambda_k}{\lambda_k}}{H\sqrt{\lambda_k-9H^2}} \]\delta_k.
\end{align}

A few remarks regarding the validity of our perturbation theory. Neglecting terms in \eqref{eq:hessDelta2} and \eqref{eq:perpVelMatrixEqDelta2} is not insensitive to the values of $\lambda_u$ and $\lambda_n$. Whenever $\lambda_u / \lambda_k \gg 1$, we naively expect from the zeroth order estimate for $\nu$ that sustained high-turn inflation is excluded. On the other hand, a consistent perturbative expansion requires the $\cO(\delta)$ correction to be parametrically smaller than the zeroth order value, i.e. it cannot offset this undesirable zeroth order value. Hence, this hierarchy of eigenvalues either fails to yield sustained inflation or falls within a regime of parameter space in which the perturbation theory cannot be trusted. Our numerical method remains self-consistent in this regime.

Utilizing the $\cO(\delta)$ results, we can solve \eqref{eq:perpMatchingDelta2} and \eqref{eq:gradMatchingDelta2} for the $\mathcal{O}\(\alpha \) \sim \cO(\delta^2)$ velocity corrections:
\begin{align}
&\begin{aligned}
\df_v^{(2)} = &\pm \frac{3 H V_v }{2 \lambda_k^3 \left(\lambda_k - 9 H^2 \right)^2} \frac{\delta_k}{\alpha} \\
&\times \Bigg[
6 H \lambda_k \left(\lambda_k - 9 H^2\right)^{3/2} \sum_n \delta_n \lambda_n \\
&\qquad \pm \delta_k \Big[
162 H^4 \lambda_u^2 - 27 H^2 \lambda_k \lambda_u \left(6 H^2 + \lambda_u\right) \\
&\qquad\qquad -2 \lambda_k^3 \left(18 H^2 + \lambda_u\right) + 18 \lambda_k^2 \left(9 H^4 + 2 H^2 \lambda_u\right) + 2 \lambda_k^4
\Big]
\Bigg]
\end{aligned}
\\
&\begin{aligned}
\df_{\perp l}^{(2)} = &\frac{V_v }{2 \lambda _k^2 \left(\lambda _k-\lambda _l\right){}^2} \frac{\delta_l}{\alpha} \\
&\times \Bigg[
6 H \lambda _k \left(\lambda _k-\lambda _l\right) \sum _n \delta _n \lambda _n \\
&\qquad \pm \frac{\delta _k}{\sqrt{\lambda _k-9 H^2}} \Big[
\lambda _k^2 \left(18 H^2 \lambda _u-\lambda _l^2+2 \lambda _l \lambda _u\right) \\
&\qquad\qquad\qquad\qquad\quad +9 H^2 \lambda _k \left(\lambda _l^2-2 \lambda _l \lambda _u-4 \lambda _u^2\right) \\
&\qquad\qquad\qquad\qquad\quad -\lambda _k^3 \left(9 H^2+2 \lambda _u\right)+18 H^2 \lambda _l \lambda _u^2+\lambda _k^4
\Big]
\Bigg]
\end{aligned}
\\
&\begin{aligned}
\df_{\perp k}^{(2)} = &\frac{1}{8 V_v \lambda _k^3 \left(\lambda _k-9 H^2\right)^{5/2}} \frac{1}{\alpha} \\
&\times \Bigg[
12 H V_v^2 \lambda_k \left(\lambda_k - 9 H^2 \right)^{3/2} \left(18 H^2 - \lambda_k \right) \delta_k \sum_n \delta_n \lambda_n \\
&\qquad \mp 36 H^2 V_v^2 \lambda_u^2 \lambda_k^2 \left(\lambda _k - 9 H^2 \right)^2 \sum_l \frac{\delta_l^2}{\left(\lambda_k - \lambda_l \right)^2} \\
&\qquad \pm V_v^2 \delta _k^2 \Big[
5832 H^6 \lambda _u^2-1296 H^4 \lambda _k \lambda _u^2-4 \lambda _k^4 \left(18 H^2+\lambda _u\right) \\
&\qquad\qquad\qquad +36 \lambda _k^3 \left(9 H^4+2 H^2 \lambda _u\right)-9 \lambda _k^2 \left(36 H^4 \lambda _u-5 H^2 \lambda _u^2\right)+4 \lambda _k^5
\Big]
\Bigg]
\end{aligned}
\end{align}
These expressions can in turn be used to further correct the turning rate. We plot the $\cO(\delta^2)$ numerical estimates of $\omega$ for a concrete example in Figure \ref{fig:helix_omega}.

\section{Examples}
\label{sec:examples}
Although we leave an extensive search though known potentials for slow-roll, rapid-turn inflation for future work, in this section we examine a few noteworthy cases.
We study the only previously published $N_f>2$ rapid-turn solutions we know of: hyperinflation and the helix, as well as search two well-studied type-IIA constructions.
In both type-IIA constructions, we fail to find long-lasting rapid-turn inflation, and are able to exclude one of them entirely. In the other, we find initial conditions that support high turning for a short period, but quickly decay.
In all examples, when appropriate, we compare and contrast the aligned hessian approximation with our numerical method.

\subsection{Hyperinflation}
\label{sec:hyperinflation}
Hyperinflation \cite{Brown:2017osf,Mizuno:2019pcm,Bjorkmo:2019aev} is a rapidly turning solution possible whenever the field space is hyperbolic, and the potential sufficiently steep around the origin.
As previously noted by \cite{Christodoulidis:2019mkj}, it is a special case of sidetracked inflation \cite{Renaux-Petel:2015mga,Garcia-Saenz:2018ifx}.
\par Under the assumptions of generalized hyperinflation \cite{Bjorkmo:2019aev}, the covariant Hessian takes a particularly simple form: 
\begin{align}\label{eq:hessGenHyper}
V_{;ab} = V_{;vv}v_a v_b + \(\frac{V_{;v}}{L} \)\perp_{ab}
\end{align}
where $L$ is the geometric scale appearing in the field space metric. I.e., in an orthonormal basis the Riemann tensor takes the form 
\begin{align}
R^a_{bcd} = -(\delta^a_c\delta_{bd} - \delta^a_d\delta_{bc})/L^2.
\label{eq:hyperinflation_riemann}
\end{align}

This model is directly amenable to the aligned Hessian approximation (and was our inspiration to study aligned Hessians in the first place). Using \eqref{eq:gradVelGradEvec}-\eqref{eq:omegaEigenvalue} and \eqref{eq:diagEtaGeneral} with $\lambda_k = V_{;v}/L$, we recover
\begin{align}
\df_v &= -3 H L \\
\df_\perp^2 &=  L V_{;v} - 9 H^2 L^2\\
\epsilon &= \frac{L V_{;v}}{2H^2} = \frac{3\epsilon_L}{2 + \epsilon_L} \\
\eta &= -3\eta_L + 2\epsilon \\
\frac{\omega^2}{H^2} &= \frac{V_{;v}}{H^2L} - 9
\end{align}
where
\begin{align}
\epsilon_L &\equiv \frac{L V_{;v}}{V}, \\
\eta_L &\equiv \frac{L V_{;vv}}{V_{;v}} 
\end{align}
in agreement with \cite{Bjorkmo:2019aev}.

\subsection{Helix}
\label{sec:helix}
This potential was first presented in \cite{Aragam:2019khr}, and is a three-field model with an analytically known high-turning background solution. The potential can be written
\begin{align}
V = \Lambda^4 \left( e^{z/R} + \Delta \left(1 - \exp\left[\frac{-(x - A \cos{z/f})^2 - (y - A \sin{z/f})^2)}{
         2 \sigma^2}\right]\right)\right)
\label{eq:helixpotential}
\end{align}
where the fields are $x,y,z$, the field space is Euclidean, and all other variables are positive constants. These authors found an analytic high-turning background solution (named the ``steady-state solution"; see Appendix A of their work) which we will not reproduce here.
The solution has constant slow-roll parameters
\begin{equation}
\begin{aligned}
\epsilon &= \Mpl^2\frac{1}{2 R^2} \frac{1}{1+A^2/f^2}\\
\epsilon_V &= \Mpl^2\frac{f^2}{2(A^2+f^2)^2}\left(\frac{A^2+f^2}{R^2}+\frac{4 A^2 f^2}{(f^2 - 6(A^2+f^2)R^2)^2}\right)\\
1+\frac{\omega^2}{9 H^2} &= \epsilon_V/\epsilon = 1+\frac{4 A^2 f^2 R^2}{(A^2+f^2)(f^2-6(A^2+f^2)R^2)^2}. \\
\eta &= \nu = \xi = 0
\end{aligned}
\label{eq:sssr}
\end{equation}
and a small $\Pi$-acceleration.

Because this attractor solution is analytically known, it gives us an opportunity to test consistency with the rapid-turn solution, as well as compare our purely numeric method with our aligned hessian approximation. The behavior of the $\delta$-expansion is quite different when the eigenvalue ratio $\lambda_u/\lambda_k$ is small than when it is large, so we present two sets of parameters: one with the ratio large over most of the surveyed space, and one with the ratio significantly smaller. These are available in Table \ref{tab:helix_parameters}.

At first glance, this model is not clearly amenable to the aligned Hessian approximation -- unlike the other models we examine, the field space is flat and the geometry cannot help align the Hessian. Furthermore, we also examine a highly featured region of the potential with large Hessian eigenvalues. We are fortunate that the extreme-turning limit is well satisfied: for parameter set 1, $\omega_\mathrm{extreme}^2/H^2 \sim 100$ on the steady state solution, while $\omega^2/H^2 \sim 3100$.

\begin{table}[h]
\centering
\begin{tabular}{l|l}
Parameter set 1 (small $\lambda_u / \lambda_k$)& Parameter set 2 (large $\lambda_u / \lambda_k$) \\ \hline
$R = 1.07 \Mpl$ & $R = 0.21 \Mpl$ \\
$\Delta = 8.43$ & $\Delta = 8.43$ \\
$A=3.47\times10^{-3} \Mpl$ & $A=3.47\times10^{-3} \Mpl$ \\
$f = 7.85\times10^{-4} \Mpl$ & $f = 1.8\times10^{-4} \Mpl$ \\
$\sigma = 8.91\times10^{-3} \Mpl$ & $\sigma = 10^{-2} \Mpl$
\end{tabular}
\caption{Values of the helix potential parameters (c.f. \eqref{eq:helixpotential}) showcased below.}
\label{tab:helix_parameters}
\end{table}

In Table \ref{tab:helix_SS_comparison}, we verify that the steady-state solution is recovered by the numerical and approximate rapid-turn solution methods. We use parameter set 1, although the results from both parameter sets are similar. On the steady-state solution, the Hessian alignment is good $\alpha\approx1.4\times10^{-3}$ and $\lambda_u/\lambda_k$ is small, so the aligned Hessian methods are relatively accurate. In each case, evolving the given initial conditions converges to the steady-state solution within a few e-folds.

\begin{table}[h]
\centering
\begin{tabular}{l|rrrrr|rrr}
\hline\hline
\textbf{} & \textbf{$\epsilon_H$} & \textbf{$\omega^2/H^2$} & \textbf{$\eta$} & \textbf{$\nu$} & \textbf{$\xi$} & \textbf{$x^\prime$} & \textbf{$y^\prime$} & \textbf{$z^\prime$} \\\hline
steady-state & 0.021 &  3193 & 0.000 & 0.000 &     0 & 0.187 & 0.072 & -0.045 \\
Numerical rapid-turn & 0.021 &  3085 & 0.021 & -5.528 &  1543 & 0.189 & 0.072 & -0.045 \\
Aligned hessian & 0.022 &  3115 & - & -191.831 & 26689 & 0.193 & 0.062 & -0.045 \\
Aligned hessian $\mathcal{O}(\delta)$ & 0.025 &  3093 & - & -37.587 & 26235 & 0.207 & 0.078 & -0.049 \\
Aligned hessian $\mathcal{O}(\delta^2)$ & 0.023 &  2942 & - & - & - & 0.195 & 0.070 & -0.046 \\\hline\hline
\end{tabular}
\caption{ Using parameter set 1, we compare various slow-roll parameters evaluated on: the helix's steady-state solution, the numerical solution, and the diagonal Hessian approximate solution. The parameters for which we have not analytically computed truncations in the $\delta$-expansion are left blank. At this point, $\epsilon_V = 7.48$, and the Hessian is closely aligned, $\alpha\approx1.4\times10^{-3}$. Our methods successfully identify the neighborhood of the steady-state solution as rapidly turning, and predict accurate initial velocities.  In more detail: $\epsilon$ is well captured by all methods, and $\eta$ is accurate to $\cO(\epsilon)$.  $\omega^2$ is only accurate to $\cO(\omega\epsilon)$, while $\nu$ is particularly poorly estimated and $\xi$ even worse. The initial velocities in the $x,y$ plane are very slightly over-estimated in the aligned Hessian approximation.}
\label{tab:helix_SS_comparison}
\end{table}

\begin{figure}[h]
\centering
\includegraphics[width=\textwidth]{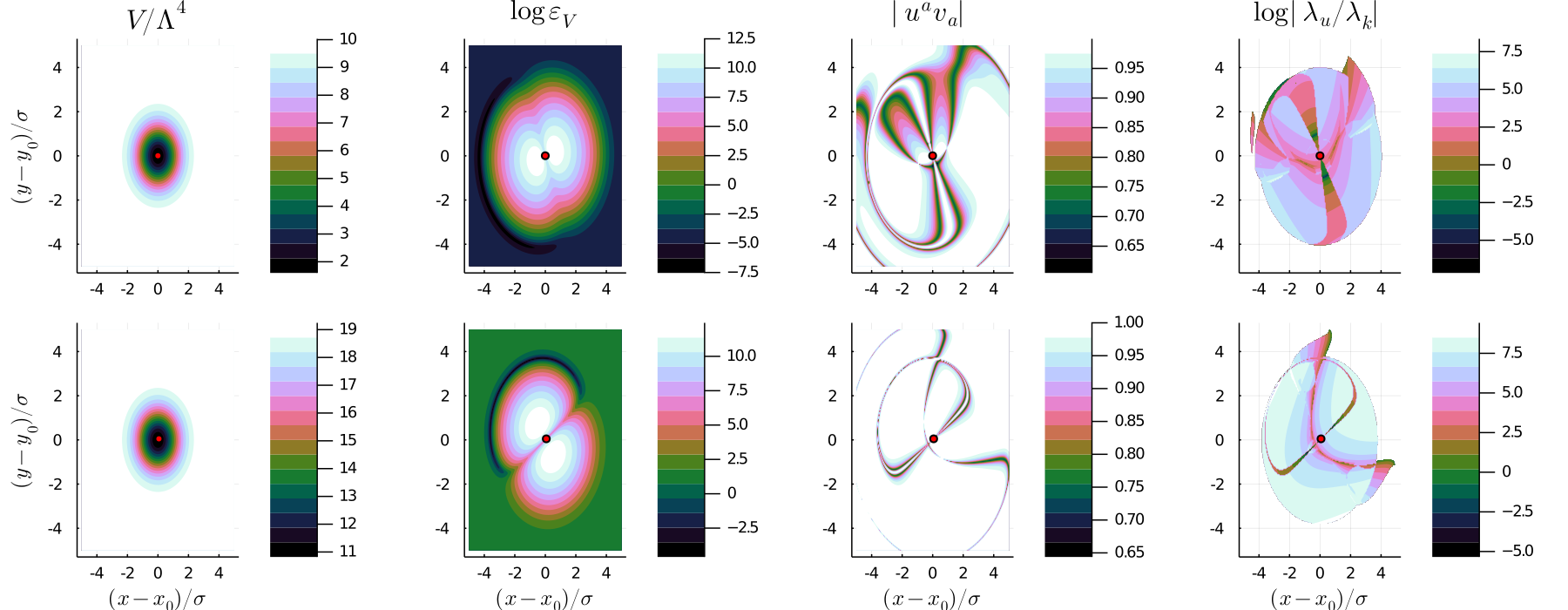}
\caption{The potential in \eqref{eq:helixpotential}, plotted over $x$ and $y$ at constant $z = 0.5 \Mpl$, centered on the helical track. The red point marks this potential's known high-turning solution. We show parameter sets 1 and 2 in the top and bottom row respectively. In the second column from the right, we show the alignment of the Hessian with the potential gradient. In the rightmost column, we show the ratio of the aligned and orthogonal eigenvalues of the Hessian: parameter set 1 has substantially lower $\lambda_u/\lambda_k$ over much of the surveyed range.}
\label{fig:helix_info}
\end{figure}

In Figure \ref{fig:helix_info}, we show the region of the potential around the center of the track plotted over $x$ and $y$, for a fixed value of $z$. Both parameter sets have $\epsilon_V < 1$ in some regions, though these are much larger for parameter set 1. The pattern of alignment (third column from the left) is similar for both parameter sets, with the regions of misalignment narrower in parameter set 2. The most notable difference between the parameter sets, though, is in the eigenvalue ratio $\lambda_u/\lambda_k$. Parameter set 2 has a much higher $\lambda_u/\lambda_k$ over the surveyed region of field space, and will invalidate our aligned Hessian results over more of the field space.

\begin{figure}[h]
\centering
\includegraphics[width=\textwidth]{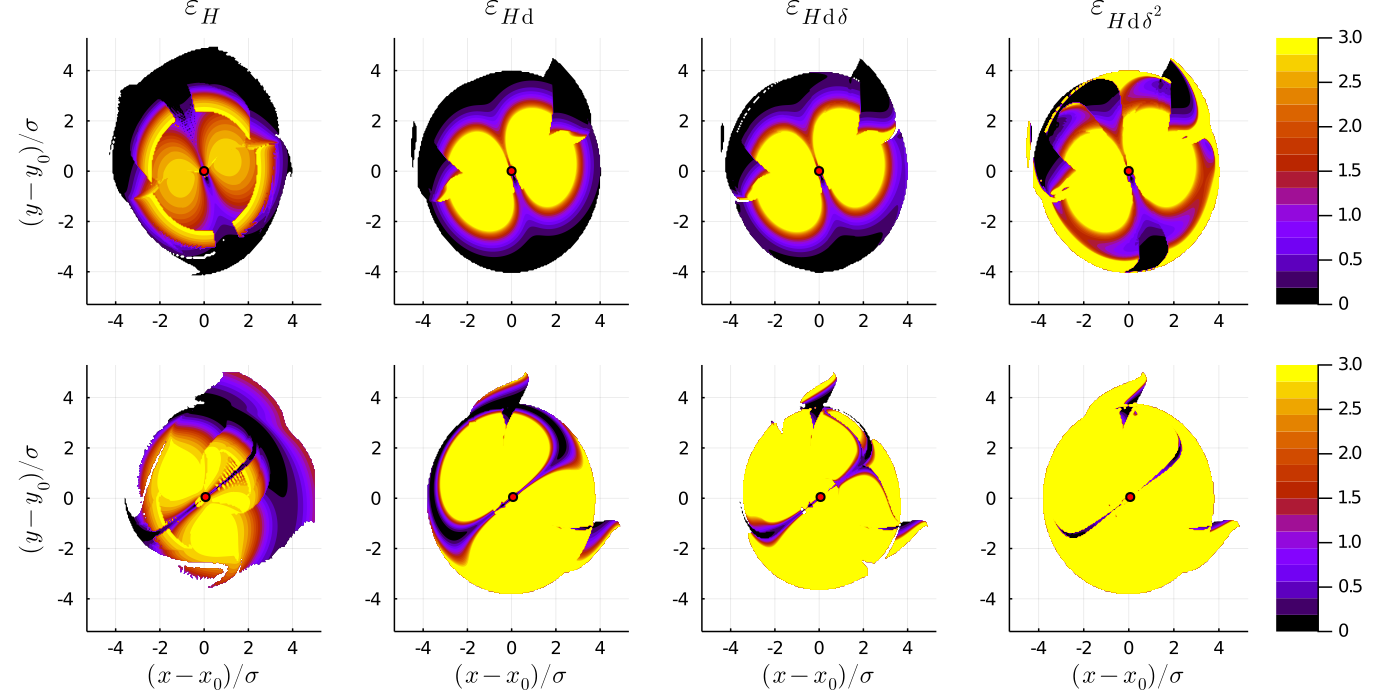}
\caption{We show the implied $\epsilon$ from the solution with the smallest $|\xi|$, for parameter sets 1 and 2 in the upper and lower rows respectively. The first column displays the numerical method calculation, and the remaining columns show increasingly higher order results from the diagonal Hessian approximation. While the diagonal method begins converging for parameter set 1, in parameter set 2 regions with a large $\lambda_u/\lambda_k$ diverge (compare with the rightmost column of \ref{fig:helix_info})}.
\label{fig:helix_eh}
\end{figure}

\begin{figure}[h]
\centering
\includegraphics[width=\textwidth]{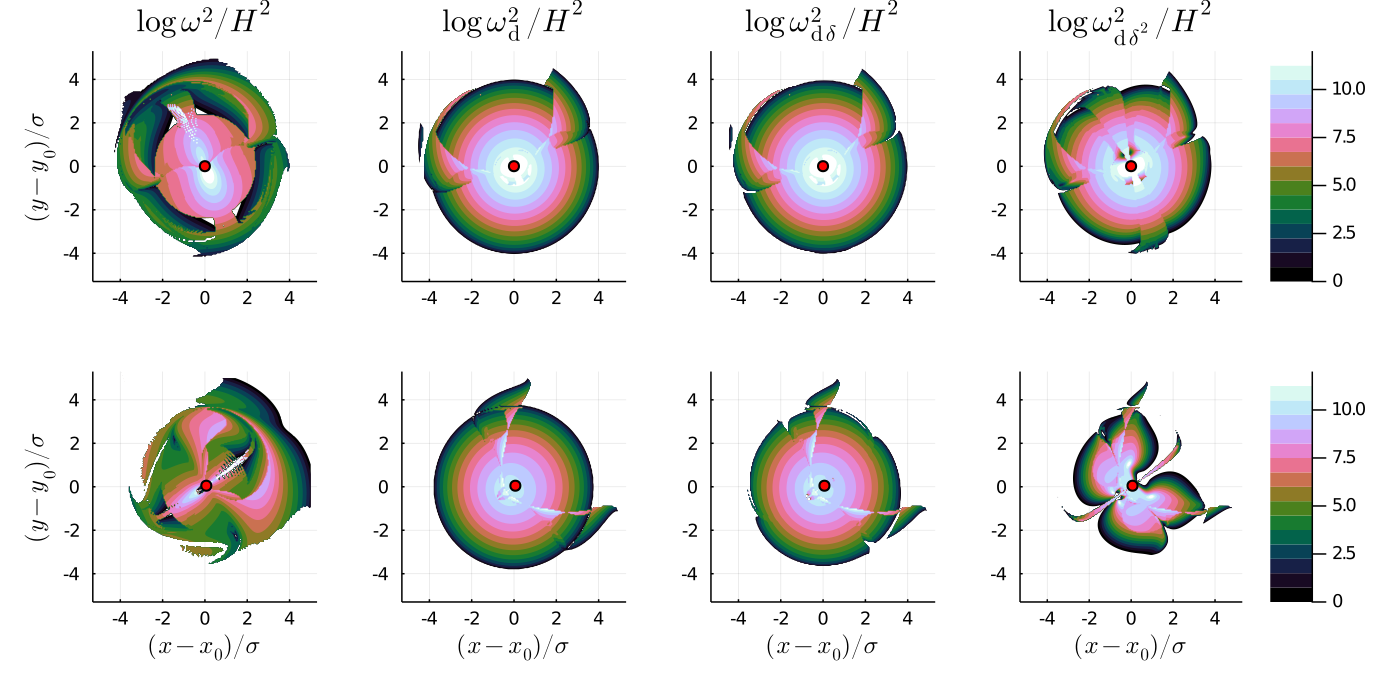}
\caption{We show the implied $\omega/H$ from the solution with the smallest $|\xi|$, for parameter sets 1 and 2 in the upper and lower rows respectively. While the diagonal method begins converging for parameter set 1, in parameter set 2 regions with a large $\lambda_u/\lambda_k$ diverge (compare with the rightmost column of \ref{fig:helix_info})}
\label{fig:helix_omega}
\end{figure}

\begin{figure}[h]
\centering
\includegraphics[width=\textwidth]{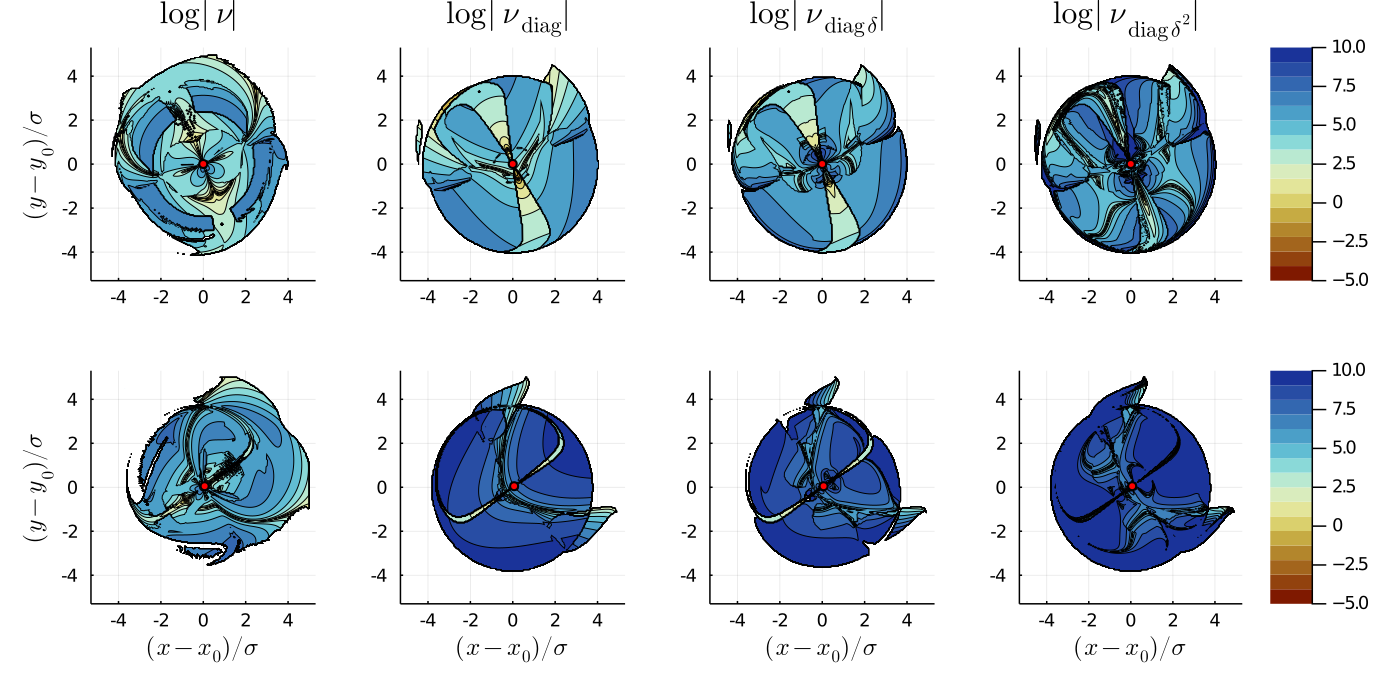}
\caption{We show the implied $\nu$ from the solution with the smallest $|\xi|$, for parameter sets 1 and 2 in the upper and lower rows respectively. While the diagonal method begins converging for parameter set 1, with parameter set 2 and its larger $\lambda_u$, most of the surveyed range is orders of magnitude misestimated. Even in the perturbatively stable regions, $\nu$ converges noticably slower than $\epsilon$ or $\omega/H$. Comparing with \eqref{eq:ehDelta}-\eqref{eq:nuDelta}, we see that corrections to $\nu$ depend on a higher power of $\lambda_u/\lambda_k$.}
\label{fig:helix_nu}
\end{figure}

In Figures \ref{fig:helix_eh}-\ref{fig:helix_nu}, we compare the predicted $\epsilon$, $\omega/H$, and $\nu$ respectively, as estimated by our numerical and aligned/diagonal Hessian results. As in the previous figure, we plot these paramters fixing $z$.
In the numerical plots (the leftmost column of each figure), we identify the possible solutions using the method of Section \ref{sec:numericalmethod}. At every point, out of the $2N_f -1 = 5$ possible solutions, we plot the solution with the smallest $|\xi|$. If the numerical method is unable to recover any solution at a point, the plot is left unfilled. For the diagonal method plots (the right three columns of each figure), to identify $\lambda_k$ and the sign of $\dot\phi_k$, we again chose the solution with the smallest implied value of $|\xi|$. In contrast to the numerical method, the slow-roll, slow-turn solution is not automatically incorporated as an alternative. The diagonal method plots, then, will never display the slow-roll, slow-turn solution as an option, even when it has the smallest $|\xi|$. Such points are rare in the surveyed parameter space of the helix, since $\epsilon_V > 1$ in the interior of the helical track (see the second-to-left column of Figure \ref{fig:helix_info}). When $\lambda_u/\lambda_k$ is large, the perturbation series in $\delta$ begins to break down. This is particularly apparent for parameter set 2, given in the bottom row of Figure \ref{fig:helix_eh}.

Overall, we approximately recover the analytically known solution, and confirm the validity of the aligned Hessian approximation when the amount of alignment and eigenvalue ratio $\lambda_u/\lambda_k$ permit the $\delta$-expansion to converge.

\FloatBarrier

\subsection{A 2-field type-IIA potential}
This potential was shown \cite{Hertzberg:2007wc} to have $\epsilon_V \geq 27/13$ everywhere, and therefore forbids slow-roll slow-turn inflation. In this section, we show that it additionally fails to support an extended period of slow-roll, rapid-turn inflation for any parameter values.

The potential can be written
\begin{align}
V = \frac{A_0}{\tau ^4 \rho^{-3}}+\frac{A_2}{\tau ^4 \rho^{-1}} + \frac{A_3}{\rho^3 \tau^2}+\frac{A_4}{\rho  \tau^4}+\frac{A_6}{\rho^3 \tau^4}+\frac{A_\mathrm{D6}}{\tau^3}-\frac{A_\mathrm{O6}}{\tau^3}
\end{align}
where the $A$ constants are positive functions of other fields in the construction (here taken to be constant), $\rho = \exp(2\hat\rho / 3)$, $\tau = \exp(\hat\tau/2)$, and the field space metric is $\delta_{ij}$ in the $(\hat\rho, \hat\tau)$ coordinates.

Because this model has only two fields, we may use the $\perp^a$-dependent expressions \eqref{eq:Bjorkmo13} and \eqref{eq:omegaHessianSimple} directly, without solving the equations of motion. As in Bjorkmo's case, there is a unique unit vector (up to a sign) $w^a$ that is perpendicular to the gradient. If there is a region of the potential with small $\epsilon_H$, large $\omega/H$, and small $\nu$, \eqref{eq:Bjorkmo13} and \eqref{eq:omegaHessianSimple} must agree. For the rest of this subsection, we name the corresponding $\omega$'s from these expressions $\omega_1$ and $\omega_2$ respectively.

\begin{figure}[h]
\centering
\includegraphics[width=0.8\textwidth]{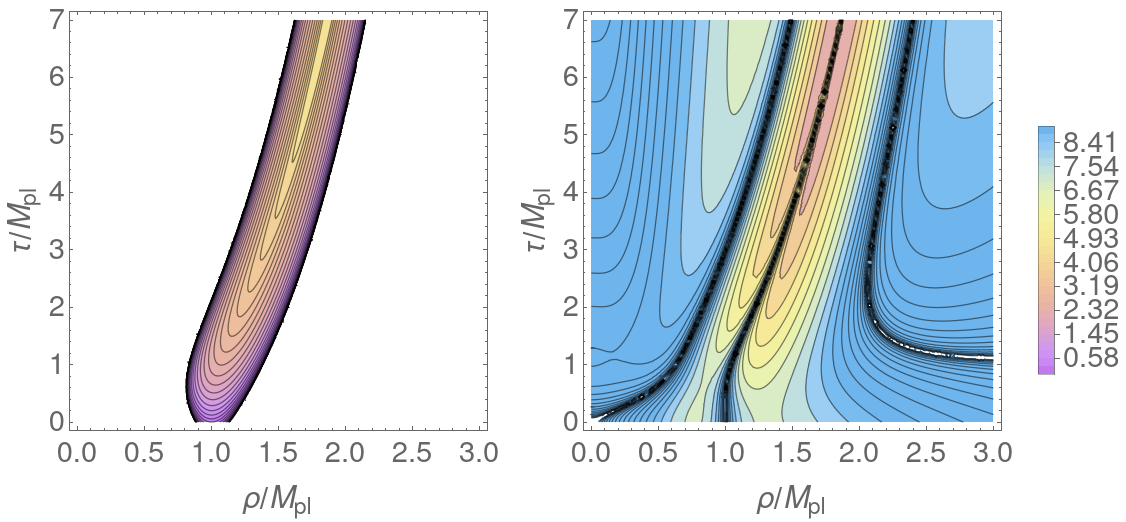}
\caption{The log of $\omega^2/H^2$, as defined in  \eqref{eq:Bjorkmo13} (left) and \eqref{eq:omegaHessianSimple} (right), for parameters $A_0 = A_2 = A_3 = A_4 = A_6 = A_\mathrm{D6} = 1,\, A_\mathrm{O6} = 3$ with $\Mpl=1$. Note that we plot with respect to the noncanonical fields $(\rho, \tau)$, not the canonical $(\hat\rho, \hat\tau)$. In regions of the potential with $\log(\omega^2/H^2) < 0$, we have left the plot colored white.}
\label{fig:twofield_omegas}
\end{figure}

In Figure \ref{fig:twofield_omegas}, we plot these two expressions for a particular value of the $A$ constants.

In order to rule out slow-roll, rapid-turn inflation, we numerically minimize a cost function of our parameters
\begin{align}
\mathrm{cost}_{N_f = 2} \equiv \frac{|\omega_1^2 - \omega_2^2|}{H^2} + B\frac{2 \epsilon_V}{2 + (\omega_1^2 + \omega_2^2)/(9H^2)}
\label{eq:cost2}
\end{align} where the second term is approximately $\epsilon$ when the two $\omega$ expressions are equal (c.f. \eqref{eq:Ev_Eh_relationship}), and $B$ trades the relative importance of the difference of the $\omega$'s and the size of $\epsilon$.
Numerically minimizing \eqref{eq:cost2} with several values of $B$ and the fields and parameters only constrained to be positive, we find, at best, an $\epsilon = 0.45$, with $\omega_1^2 = 65.8 H^2$, $\omega_2^2 = 66.8 H^2$. This is a short-lived solution with $\eta = 0.9 \sim 2 \epsilon$.

As a check on our diagonal method, we also use it to analyze this potential, and rule out some regions of field space.
We denote the eigenvectors of $V_{;ab}$ as $\{ u^a, t^a\}$, where $u^a$ is chosen to be the eigenvector most aligned with the gradient. Then, per \eqref{eq:ehEigenvalue}, we would need $V_{tt} \gg 9 H^2 \epsilon_V$ along an inflationary trajectory in order to support inflation with a small $\epsilon$. By taking limits of the eigenvalues and eigenvectors, we can see that this condition is violated in asymptotic field space ($\hat\rho,\hat\tau \rightarrow \pm \infty$), where $V_{tt}/H^2\rightarrow 0$, $u^a \rightarrow v^a$, and $\epsilon_V\rightarrow (4,7)$ exponentially (both values of $\epsilon_V$ are possible depending on the direction of the limit). Then the only region that could possibly support inflation is near the origin.

An extensive search using our numerical method over the region $A_\circ \in [0,10]$, $\rho,\tau \in [0,10\Mpl]$ produces at minimum, a cost function (c.f. \eqref{eq:cost}) of approximately $0.03$, corresponding to $\epsilon\sim 0.49$, $\eta\sim0.49$, $\xi\sim2.6$, and $\nu \sim -0.05$. Evolving the point gives roughly $0.35$ e-folds of inflation. This solution is at a different point than the purely two-field method above, but of comparable quality. 

To our knowledge, this constitutes the first potential and metric excluded from supporting slow-roll, rapid-turn inflation. In this case, the purely two-field methods or the diagonal and numerical method prove equally powerful. The diagonal Hessian method immediately excludes asymptotic field space, while numerically searching with either the two-field or the full numeric method excludes regions around the origin. It is our hope that a sizable variety of potentials and metrics, even those with significantly more fields, will be amenable to a similar procedure.

\subsection{An 8-field type-IIA potential (DGKT)}
This potential, first presented in \cite{DeWolfe:2005uu} and examined in Appendix A of \cite{Hertzberg:2007ke}, is from the K\"ahler sector of a type-IIA compactification. The K\"ahler potential is
\begin{align}
K=-\log(32 \, b_1 b_2 b_3 b_4^4),
\end{align}
where the real-valued fields can be labelled $\phi^I = \{a_1,a_2,a_3,a_4,b_1,b_2,b_3,b_4\}$. The field-space metric is
\begin{align}
G_{IJ} =  \frac{1}{2} \, \mathrm{diag}\left(\frac{1}{b_1^2},\frac{1}{b_2^2},\frac{1}{b_3^2},\frac{4}{b_4^2} ,\frac{1}{b_1^2} ,\frac{1}{b_2^2},\frac{1}{b_3^2},\frac{4}{b_4^2} \right).
\end{align}
Although the field space does not fit the totally isotropic form necessary for hyperinflation \eqref{eq:hyperinflation_riemann}, the scalar curvature is constant and negative $R=-13$.

The potential is known up to an overall constant $V_\textrm{flux}$ (irrelevant for the background evolution) and a sign, $\delta=\pm 1$
\begin{equation}
\begin{aligned}
V = \frac{V_\textrm{flux}}{32 b_1 b_2 b_3 b_4^4} \bigg[& -4 a_1 a_2 a_3 \delta  \left(a_1+a_2+a_3+2 \sqrt{2} a_4 \right)-4 \delta (a_2 a_3 b_1^2 + a_1 a_3 b_2^2 + a_1 a_2 b_3^2) \\
& + 2 ( a_2^2 a_3^2 b_1^2 + a_1^2 a_3^2 b_2^2 + a_1^2 a_2^2 b_3^2 ) + 2(a_1^2 b_2^2 b_3^2 + a_2^2 b_1^2 b_3^2 + a_3^2 b_1^2 b_2^2) + 2 a_1^2 a2^2 a_3^2 \\
& + 2 \left( a_1 + a_2 + a_3 +2 \sqrt{2} a_4 \right)^2-8 \sqrt{2} b_1 b_2 b_3 b_4 \\
&+ 2 b_1^2 b_2^2 b_3^2 + 2( b_1^2 + b_2^2 + b_3^2) + 4 b_4^2 \bigg].
\end{aligned}
\end{equation}

This potential contains several points that at first glance appear to support a rapid-turn solution.
With $\delta=-1$, one such point is $\phi_\star^a \equiv \{18.77, -0.0517, -0.0264, -6.142, 0.00174, 5.968\times 10^{-5}, 4.781, 0.00254\} \Mpl$. The eigenvalues of the Hessian and the implied aligned Hessian approximation $\epsilon$ and $\omega/H$ are available in Table \ref{tab:DGKT_smalleh}. We also note that the 6-volume $\mathrm{Vol}_6 = b_1 b_2 b_3 \sim 10^{-7}$ is unphysically small, and the implied $g_s = b_4^{-1}\sqrt{\mathrm{Vol_6}/2}$ is an acceptable value of roughly $0.2$ \footnote{Similar points exist at higher (still unphysical) values of $\mathrm{Vol_6}$ as well. One such point is $\phi^a = \{ -20.0, 0.047,0.0243,1.09,3.09,0.060,5.0,1.63\}\Mpl$, with Vol$_6 \sim 0.93$, $g_s\sim 0.42$, and $\epsilon\sim 0.1$, $\lambda_k \sim 425H^2$.}.

\begin{table}
\centering
\begin{tabular}{rrrr}
\hline\hline
\textbf{$t_i^a v_a$} & \textbf{$\lambda_i/H^2$} & \textbf{$\epsilon_{H \mathrm{ d}}$} & \textbf{$\omega_{\mathrm{ d}}^2/H^2$} \\\hline
-0.000444 & -6.000067 & -10.437980 & -15.000067 \\
0.000015 & -5.999987 & -10.438120 & -14.999987 \\
-0.000117 & -5.998909 & -10.439994 & -14.998909 \\
0.000005 & 11.996639 & 5.220510 & 2.996639 \\
0.000000 & 12.000013 & 5.219043 & 3.000013 \\
-0.000023 & 12.001564 & 5.218368 & 3.001564 \\
0.999995 & 53.752597 & 1.165127 & 44.752597 \\
0.002989 & 7525.069161 & 0.008323 & 7516.069161 \\\hline
- & - & 0.008301 & 7494.410042 \\\hline\hline
\end{tabular}
\caption{The Hessian's eigenvector alignment and eigenvalues at the point $\phi_\star^a$ along with each eigenvalue's implied $\epsilon$ and $\omega/H$, assuming that eigenvalue is the $\lambda_k$ of the $\cO(\delta^0)$ aligned Hessian approximation (see \eqref{eq:ehEigenvalue}-\eqref{eq:omegaEigenvalue}). Although these expressions are invalid when $\lambda_k = \lambda_u$ or $\lambda_k < 9 H^2$, in this table we compute them anyway. The eigenvector and eigenvalue in the seventh row are the most aligned with the gradient direction, while the eigenvalue in the eighth row appears to support a rapid-turn solution. In the last row, we give the numerical method's $\epsilon$ and $\omega/H$ evaluated at the same point. For reference, at this point $\epsilon_V \approx 6.95$ and $\omega_\mathrm{extreme}/H \approx 2.17$.}
\label{tab:DGKT_smalleh}
\end{table}

Unfortunately, these initial conditions do not bear out an extended period of inflation. The slow-roll parameters $\nu$ and $\xi$ grow with higher order corrections in the $\delta$-expansion, indicating a breakdown of one of our assumptions. The full numerical method gives the reason why: evaluated at this point using \eqref{eq:omegaDefn}, it agrees well with the diagonal method's $\epsilon$ and $\omega/H$. Unfortunately, the two small-$\nu$ estimates for $\omega$ greatly differ: $\omega^2/H^2 = \{56, 7485\}$ for \eqref{eq:omegaHessianSimple} and \eqref{eq:Bjorkmo13} respectively. The measured $\nu\sim 0.25$ is not $\cO(\epsilon)$, so neglecting the accelerations while solving the equations of motion is inconsistent -- we do not have a long-lasting solution.

\begin{figure}[h]
\centering
\includegraphics[width=\textwidth]{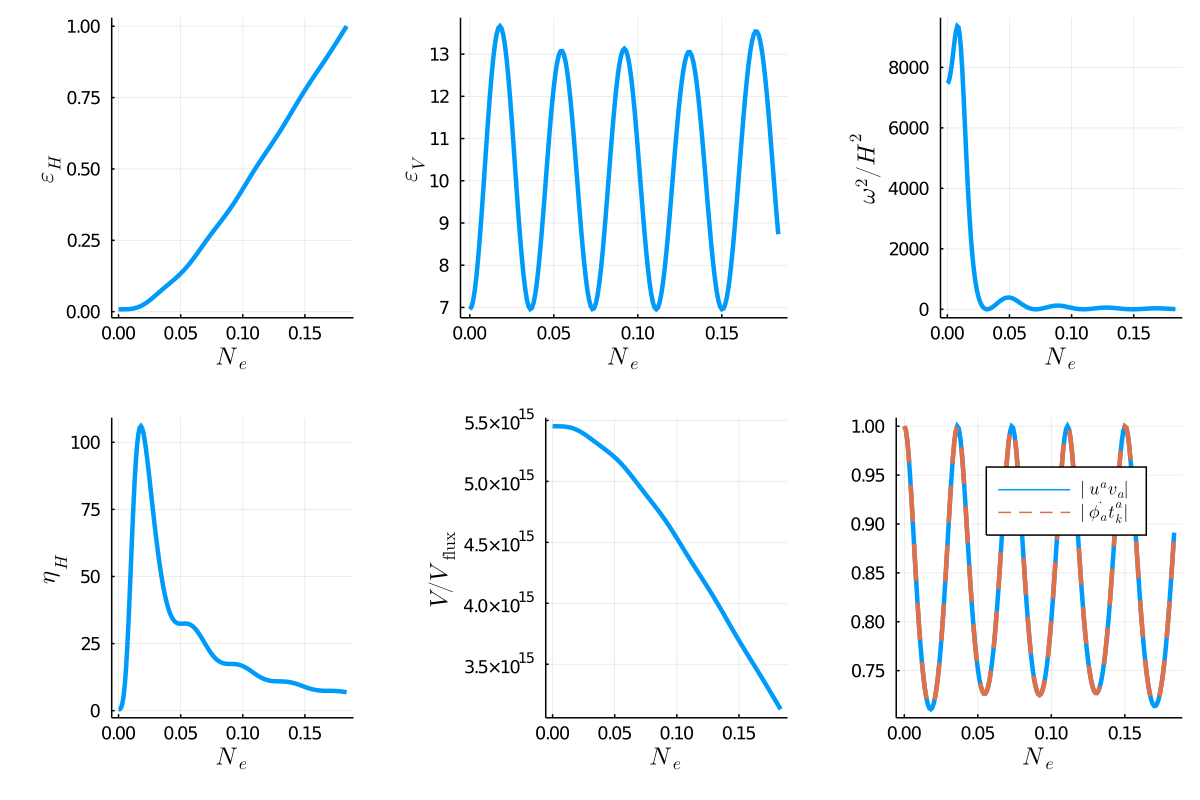}
\caption{Several slow-roll parameters and the potential numerically evolved along the trajectory beginning at $\phi_\star^a$. Though the trajectory is initially rapid-turn, $\omega/H$ rapidly decays and $\eta$ sharply rises, quickly ending inflation after $0.183$ e-folds. The rapid-turn behavior ends quickly as the Hessian and trajectory misalign within a few hundredths of an e-fold.}
\label{fig:DGKT_slowroll}
\end{figure}

\begin{figure}[h]
\centering
\includegraphics[width=\textwidth]{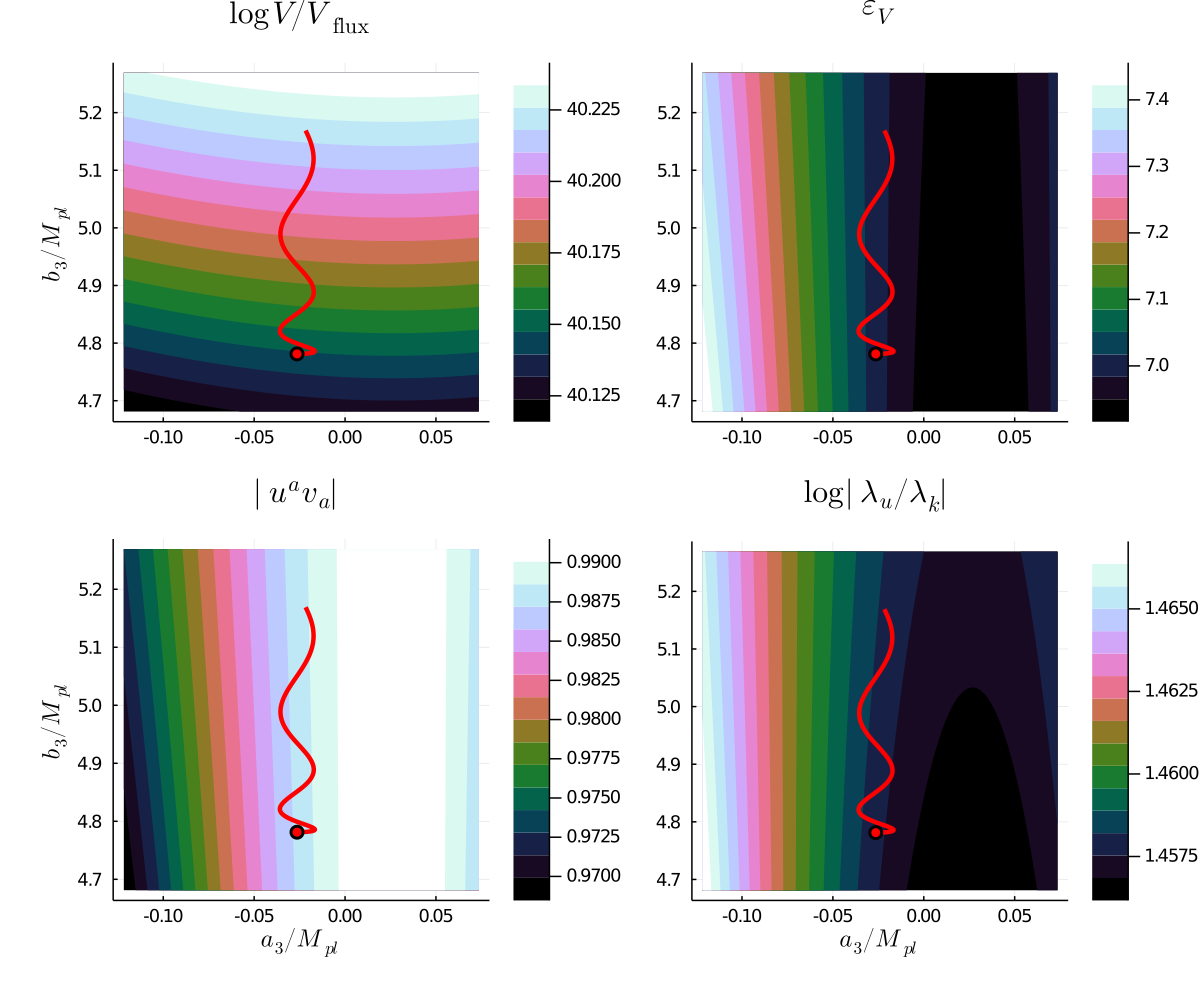}
\caption{The neighborhood of the point $\phi_\star^a$  (red circle) in the DGKT potential, with the numerically evolved trajectory in red. Though the field space is 8-dimensional, we plot the potential, Hessian alignment, $\epsilon_V$, and eigenvalue ratio sliced along the $(a_3,b_3)$ subspace. Note that, except for the initial point, the trajectory is slightly out of the plotted plane in the 6 orthogonal directions -- for accurate information about these quantities on-trajectory, see Figure \ref{fig:DGKT_slowroll}.}
\label{fig:DGKT_neighborhood}
\end{figure}

In Figure \ref{fig:DGKT_slowroll}, we evolve the $\cO(\delta^0)$ initial conditions numerically, generating $\sim 0.18$ e-folds of inflation\footnote{Note that considerably more e-folds $N_e \lesssim 5$ are possible in this potential when starting trajectories with zero initial velocities and high accelerations. These short-lived initial conditions are not rapid-turn.}. The rapid-turn initial conditions are quickly spoiled by the large increase in $\eta$ and decay in $\omega/H$.
In Figure \ref{fig:DGKT_neighborhood}, we plot the potential and several other quantities in the neighborhood of the trajectory. Note that we only slice the 8-dimensional field space in the $(a_3,b_3)$ subspace, which features the majority (but not all) of the fields' motion. As a consequence, the on-trajectory values of $\epsilon_V$ differ from the ones in the contour plot.

Other than this trajectory, an extensive search with the numerical method over $a_i=[-20,+20]\Mpl$, $b_i = [10^{-5},5]\Mpl$ found no improvement, with a minimum cost function of approximately $9.9$ (corresponding to $\epsilon \sim 0.25$, $\omega/H \sim 0.05$, $\eta \sim 20$).

We have also searched an $N_f = 4$ simplification of this potential and metric (previously studied as expression (39) of \cite{Hertzberg:2007ke}), with $a_1=a_2=a_3$ and $b_1=b_2=b_3$. No points with a Hessian like the one presented in Table $\eqref{tab:DGKT_smalleh}$ were found. The minimum cost function found was comparable to the 8-field case.

\FloatBarrier

\section{Summary and Conclusions}

We have sought to address a broad question: given a multi-field potential and field space geometry, what regions of field space admit inflationary solutions? The limited viability of slow-roll, slow-turn trajectories in generic potentials has directed our attention toward identifying rapid-turn solutions. By generalizing a previous two-field rapid-turn attractor to many fields, we identify and enumerate the conditions required for sustained rapid-turn inflation at any point in field space. In particular, we find an \textit{extreme turning} limit in which most of the accelerations are negligible. This limit is characterized by components of the potential's covariant Hessian matrix that are not present in the two-field attractor. We then apply these conditions to the equations of motion and formulate a matrix equation that constrains the field velocities. We propose a numerical algorithm to solve this equation, as well as an analytic, perturbative approach that holds when the potential gradient is strongly aligned with an eigenvector of the Hessian matrix.

Lastly, we examine a variety of EFT and Type-IIA constructions using the two approaches, comparing to known attractor solutions when applicable. In one construction, we are able to exclude slow-roll, rapid-turn inflation from the entire field space. In others, we can rule out vast regions of field space, but do not find any previously unknown long-lasting rapid-turn trajectories. With our methods, determining whether a potential and metric admit or exclude rapid-turn trajectories is now possible.

\section{Acknowledgments}

V.A. thanks Aaron Zimmerman for insightful discussions on solving differential equations. S.P. thanks the members of the Berkeley Center for Theoretical Physics for their hospitality.  The National Science Foundation supported this work under grant number PHY–1914679.

\bibliography{refs}{}

\bibliographystyle{JHEP}

\end{document}